\documentclass[aps, pre, a4paper, floatfix,  twocolumn, longbibliography, nofootinbib]{revtex4-1}

\usepackage[utf8]{inputenc}
\usepackage{epsfig}
\usepackage[T1]{fontenc}
\usepackage[english]{babel}
\usepackage[table]{xcolor}
\usepackage{t1enc}
\usepackage{graphicx}
\usepackage{amssymb}
\usepackage{amsmath}
\usepackage{relsize}
\usepackage[percent]{overpic}
\usepackage{bm}
\usepackage{hyperref}

\usepackage[normalem]{ulem}
\usepackage{grffile}
\usepackage{verbatim}
\usepackage{overpic}
\usepackage{booktabs}

\usepackage{siunitx}
\usepackage{multirow}

\newcommand{\avg}[1]{{\left<#1\right>}}

\newcommand{\dd}{\mathrm{d}}
\newcommand{\e}{\mathrm{e}}
\newcommand{\A}{\bm{A}}
\newcommand{\G}{\bm{G}}
\newcommand{\bb}{\bm{b}}

\usepackage{mathtools}
\def\multiset#1#2{\ensuremath{\left(\kern-.3em\left(\genfrac{}{}{0pt}{}{#1}{#2}\right)\kern-.3em\right)}}

\begin{document}

\title{Latent Poisson models for networks with heterogeneous density}

\author{Tiago P. Peixoto}
\email{peixotot@ceu.edu}
\affiliation{Department of Network and Data Science, Central European University, H-1051 Budapest, Hungary}
\affiliation{ISI Foundation, Via Chisola 5, 10126 Torino, Italy}
\affiliation{Department of Mathematical Sciences, University of Bath, Claverton Down, Bath BA2
  7AY, United Kingdom}
\begin{abstract}
  Empirical networks are often globally sparse, with a small average
  number of connections per node, when compared to the total size of the
  network. However, this sparsity tends not to be homogeneous, and
  networks can also be locally dense, for example with a few nodes
  connecting to a large fraction of the rest of the network, or with
  small groups of nodes with a large probability of connections between
  them. Here we show how latent Poisson models that generate hidden
  multigraphs can be effective at capturing this density heterogeneity,
  while being more tractable mathematically than some of the
  alternatives that model simple graphs directly. We show how these
  latent multigraphs can be reconstructed from data on simple graphs,
  and how this allows us to disentangle disassortative degree-degree
  correlations from the constraints of imposed degree sequences, and to
  improve the identification of community structure in empirically
  relevant scenarios.
\end{abstract}

\maketitle

\section{Introduction}

One of the most important properties of empirical networks ---
representing the pairwise interactions of social, biological,
informational and technological systems --- is that they exhibit a
strong structural heterogeneity, while being globally
sparse~\cite{newman_networks:_2010}. The latter property means that most
possible connections between nodes are not observed, which as a
consequence means that, on average, the probability of observing a
connection between two nodes is very small, and hence the typical number
of connections each node receives is much smaller than the total number
of nodes in the network. For example, even though the global human
population is in the order of billions, most people interact only with a
far smaller number of other people. Nevertheless, such network systems
are rarely \emph{homogeneously} sparse: instead, local portions of the
network can vary greatly in their number of interactions. As has been
widely observed~\cite{barabasi_emergence_1999}, the number of neighbors
of each node is very often broadly distributed, typically spanning
several orders of magnitude. In addition, networks exhibit diverse kinds
of mixing patterns in relation to the degrees~\cite{newman_mixing_2003},
e.g. nodes may connect to other nodes with similar degree
(assortativity), or nodes with high degree may connect preferentially
with nodes of low degree and vice-versa (disassortativity). It is possible
also for networks to possess communities of tightly connected
nodes~\cite{fortunato_community_2010}, such that the probability of a
link existing between members of these subgroups far exceeds the global
average. The existence of such heterogeneous mixing patterns serves as a
signature of the process responsible for the network formation and may
give insight into its functional aspects.

A central complicating factor in the characterization and understanding
of the different kinds of mixing patterns in networks is that they
cannot be fully understood in isolation. For example, although networks
with degree heterogeneity can exhibit in principle any kind of mixing
pattern, there is a stronger tendency of very heterogeneous networks to
exhibit degree
disassortativity~\cite{park_origin_2003,johnson_entropic_2010,del_genio_all_2011}. This
is because once the degrees of a fraction of the nodes become comparable
to the total number of nodes in the network, there is no other option
than to connect them with nodes of lower degree. Since it is not
possible to fully decouple degree heterogeneity from mixing, it can
become difficult to determine whether the latter is simply a byproduct
of the former, or if it can be related to other properties of network
formation.

The degree disassortativity induced from broad degree distributions can
also occur in networks exhibiting community structure, and in a similar
way: if a node has a degree comparable to the number of nodes in the
community to which it belongs, it will tend to be connected to nodes of
the same community with a smaller degree. The resulting mixing pattern
may confuse community detection methods that do not account for this
possibility, which will mistake the pattern that arises from a purely
intrinsic constraint, with one that needs an extrinsic explanation in
the form of a different division of the network into groups.

In this work we address the problem of describing degree and density
heterogeneity by considering models of random multigraphs, i.e. where
more than one link between nodes is allowed, as well as self-loops,
following a Poisson distribution, where a full decoupling of the degree
distribution and degree mixing patterns is in fact possible. These can
be transformed into models of simple graphs by erasing self-loops and
collapsing any existing multiedges into a single edges. Conversely, we
can recover the decoupling of degree variability and mixing by
\emph{reconstructing} an underlying multigraph from a given observation
of a simple graph. Then, by inspecting the inferred multigraph, we can
finally determine what is cause and byproduct.

We also show how latent Poisson models can be employed in the task of
community detection in the presence of density and degree
heterogeneity. When dealing with simple graphs, degree
correction~\cite{karrer_stochastic_2011}, as we show, is in general is
not sufficient to disentangle community structure from induced degree
disassortativity. When performing latent multigraph reconstruction, we
demonstrate that this becomes finally possible. Furthermore, we show how
the latent multigraph approach is more effective at describing the
heterogeneous density of many networks, when compared to just using a
multigraph model directly to represent a simple graph, as is often
done~\cite{peixoto_bayesian_2019}. In particular, we show how this
increases our ability to uncover smaller groups in large networks.

This paper is divided as follows. We begin in Sec.~\ref{sec:maxent} by
formalizing the intrinsic effect of degree constraints by employing the
maximum-entropy principle, and we show how the Poisson model appears
naturally when multiedge distinguishability is taken into
consideration. In Sec.~\ref{sec:erased} we describe the
erased Poisson model for simple graphs, we compare its induced degree
correlations with alternative maximum-entropy models, and present
methods of Bayesian inference capable of reconstructing it from simple
graph data. We then show how it can shed light into the origins of
degree disassortativity in empirical networks. In Sec.~\ref{sec:community}
we show how the erased Poisson model can improve the task of community
detection, by allowing the induced degree mixing to be decoupled from
the modular network structure, and also describe arbitrary density
heterogeneity, and thereby enhance the resolution of small dense
communities in large globally sparse networks. We finalize in
Sec.~\ref{sec:conclusion} with a conclusion.

\section{Maximum-entropy ensembles for simple and multigraphs}\label{sec:maxent}

One of our primary objectives is to model the effects of degree
heterogeneity in network structure. To this end, we will concern
ourselves with network ensembles that satisfy the constraint that the
expected degrees of the nodes are given as parameters.  Specifically, if
$P(\A)$ is the probability that network $\A$ occurs in the ensemble, we
have that the following condition needs to hold
\begin{equation}\label{eq:const_k}
  \sum_{\A} P(\A) \sum_jA_{ji} = \hat{k}_i,
\end{equation}
for a given expected degree sequence $\bm{\hat k} = \{\hat k_1,\dots,
\hat k_N\}$, where $A_{ij}$ determines the number of edges between nodes
$i$ and $j$, $k_i=\sum_jA_{ji}$ is the degree of node $i$, and and $N$
is the number of nodes in the network. Given the constraints of
Eq.~\ref{eq:const_k}, there are many choices of ensemble $P(\A)$ that
satisfy it. Since we want to understand the intrinsic effect of the
imposed degrees on other aspects of the network structure, we are
interested in the choice of $P(\A)$ that is maximally uniform, or
agnostic, with respect to the possible networks, conditioned only that
the above constraint is satisfied. More formally, this means we want the
choice that maximizes the ensemble entropy~\cite{jaynes_probability_2003,bianconi_entropy_2009}
\begin{equation}
  \mathcal{S} = -\sum_{\A}P(\A)\ln P(\A).
\end{equation}
Employing the method of Lagrange multipliers to perform the constrained
maximization yields a product of independent distributions for each
entry in the adjacency matrix,
\begin{equation}\label{eq:max_ent_adj}
  P(\A) = \prod_{i<j}\frac{(\theta_i\theta_j)^{A_{ij}}}{Z_{ij}},
\end{equation}
where the $\bm\theta$ are ``fugacities'' (exponentials of Lagrange
multipliers) that keep the constraints in
place~\cite{park_origin_2003,bianconi_entropy_2009,anand_gibbs_2010,squartini_breaking_2015},
and $Z_{ij}=\sum_{A_{ij}}(\theta_i\theta_j)^{A_{ij}}$ is a normalization
constant, comprised of a sum over all possible values of $A_{ij}$. Thus,
the value of $Z_{ij}$ will be different depending on whether we are
dealing with simple graphs or multigraphs. For the case of simple graphs
with $A_{ij}\in\{0,1\}$, we have $Z_{ij} = 1+\theta_i\theta_j$, which
results in independent Bernoulli distributions for every node pair,
\begin{equation}\label{eq:maxent_simple}
  P(\A) = \prod_{i<j}\frac{(\theta_i\theta_j)^{A_{ij}}}{1+\theta_i\theta_j},
\end{equation}
with mean values
\begin{equation}\label{eq:avga_simple}
  \avg{A_{ij}} = \frac{\theta_i\theta_j}{1+\theta_i\theta_j}.
\end{equation}
In order for the constraints of Eq.~\ref{eq:const_k} to be fulfilled,
the fugacities need to be chosen by solving the system of nonlinear
equations
\begin{equation}
  \sum_{j\ne i} \frac{\theta_i\theta_j}{1+\theta_i\theta_j} = \hat{k}_i,
\end{equation}
which in general does not admit closed analytical solutions, and needs
to be solved numerically.

For multigraphs with $A_{ij}\in\mathbb{N}_0$, we have instead
$Z_{ij}=\sum_{A_{ij}=0}^{\infty}(\theta_i\theta_j)^{A_{ij}}=(1-\theta_i\theta_j)^{-1}$,
assuming $\theta_i\theta_j < 1$, which results in a product of geometric
distributions,
\begin{equation}
  P(\A) = \prod_{i<j}(\theta_i\theta_j)^{A_{ij}}(1-\theta_i\theta_j),
\end{equation}
with mean values
\begin{equation}\label{eq:avga_multi}
  \avg{A_{ij}} = \frac{\theta_i\theta_j}{1-\theta_i\theta_j},
\end{equation}
and the fugacities are obtained by solving an analogous but different
system of equations
\begin{equation}
  \sum_{j\ne i} \frac{\theta_i\theta_j}{1-\theta_i\theta_j} = \hat{k}_i,
\end{equation}
which also cannot be solved in closed form in general.

In both of the above cases, if all imposed degrees $\hat k_i$ are
sufficiently smaller than $\sqrt{2E}$, with $2E=\sum_i\hat{k}_i$ being
twice the number of expected edges, then in the limit $N\gg 1$ the
fugacities can be obtained approximately as
\begin{equation}
  \theta_i \approx \frac{\hat{k}_i}{\sqrt{2E}}.
\end{equation}
In this case the expected value of the adjacency matrix becomes
\begin{equation}
  \avg{A_{ij}} \approx \theta_i\theta_j = \frac{\hat{k}_i\hat{k}_j}{2E},
\end{equation}
both for simple and multigraphs, and hence the difference between those
ensembles vanish. In this situation, the expected number of edges
between nodes is in the order of $1/N$, for sparse networks with $E \sim
O(N)$, and thus the networks are also locally sparse, since no portion
of the network is connected with high probability. The ensemble does not
possess intrinsic degree correlations between neighbors, since the
expected value of the adjacency matrix is simply the product of the
fugacities.  However, if the expected degrees $\bm{\hat k}$ are broadly
distributed, with a fraction of them approaching or exceeding
$\sqrt{2E}$ (known as the ``structural
cut-off''~\cite{boguna_cut-offs_2004}), this assumption will no longer
hold, even if the network is globally sparse with $E \sim O(N)$. In this
situation, typical networks sampled from the ensemble will exhibit
intrinsic nontrivial mixing patterns. We will return to this in
Sec.~\ref{sec:correlations}, but for now we move to maximum-entropy
ensembles with distinguishable multiedges.

\subsection{Distinguishable multiedges and the Poisson model}

We now consider a third situation when multiple edges between nodes are
allowed, but the individual edges between the same two nodes can be
distinguished from one another. We do so by allowing the edges to belong
to one of $M$ discrete types. We implement this by introducing a binary
variable $X_{ij}^m \in \{0,1\}$ specifying whether an edge of type $m
\in [1, M]$ exists between nodes $i$ and $j$, such that the adjacency
matrix of the associated multigraph becomes $A_{ij}=\sum_{m=1}^M
X_{ij}^m$, so that $A_{ij} \in [0, M]$.  By maximizing the entropy of
this augmented ensemble while imposing the same degree constraints of
Eq.~\ref{eq:const_k}, we obtain an equation similar in form to
Eq.~\ref{eq:max_ent_adj},
\begin{equation}
  P(\bm X) = \prod_{i<j}\frac{(\theta_i\theta_j)^{\sum_mX^m_{ij}}}{Z_{ij}},
\end{equation}
but with a different normalization
\begin{equation}
  Z_{ij} = \sum_{A_{ij}=0}^M{M\choose A_{ij}}(\theta_i\theta_j)^{A_{ij}} = (1+\theta_i\theta_j)^M.
\end{equation}
If we now consider the associated multigraph ensemble, by ignoring the
edge types, we obtain a product of binomial distributions
\begin{align}
  P(\bm A) &= \sum_{\bm X}P(\bm X)\prod_{i<j}\delta_{A_{ij,\sum_mX_{ij}^m}}\\
  &= \prod_{i<j}{M \choose A_{ij}}\left(\frac{\theta_i\theta_j}{1+\theta_i\theta_j}\right)^{A_{ij}}
  \left(1-\frac{\theta_i\theta_j}{1+\theta_i\theta_j}\right)^{M-A_{ij}}.
\end{align}
Taking the limit $M\to\infty$, and making the variable transformation
$\theta_i\to\theta_i/\sqrt{M}$, while keeping the constraints of
Eq.~\ref{eq:const_k} fixed, we obtain a product of Poisson distributions
\begin{equation}\label{eq:poisson}
  P(\bm A) = \prod_{i<j}\frac{(\theta_i\theta_j)^{A_{ij}}\e^{-\theta_i\theta_j}}{A_{ij}!}.
\end{equation}
In this case the degree constraints take a simpler form
\begin{equation}
  \hat{k}_i = \lim_{M\to\infty} \sum_{j\ne i}\frac{\theta_i\theta_j}{1+\frac{\theta_i\theta_j}{M}} = \theta_i\sum_{j\ne i}\theta_j.
\end{equation}
The above model becomes even more convenient if we allow for self-loops,
i.e. $A_{ii} > 0$. Repeating the same calculations we obtain
\begin{equation}\label{eq:poisson-sl}
  P(\bm A) = \prod_{i<j}\frac{(\theta_i\theta_j)^{A_{ij}}\e^{-\theta_i\theta_j}}{A_{ij}!}\prod_i\frac{(\theta_i^2/2)^{A_{ii}/2}\e^{-\theta_i^2/2}}{(A_{ii}/2)!},
\end{equation}
where we adopt the convention that $A_{ii}$ is twice the number of
self-loops incident on node $i$. With this simple modification the
degree constraints now become
\begin{equation}
  \hat{k}_i = \theta_i\sum_j\theta_j.
\end{equation}
Unlike any of the previous models considered, the above equations can be
directly solved as
\begin{equation}
  \theta_i = \frac{\hat{k}_i}{\sqrt{2E}},
\end{equation}
once more with $2E = \sum_i \hat{k}_i$. The mean of the adjacency entry
is then
\begin{equation}
  \avg{A_{ij}} = \theta_i\theta_j = \frac{\hat{k}_i\hat{k}_j}{\sqrt{2E}}.
\end{equation}
This model, therefore, becomes asymptotically equivalent to the simple
and multigraph ensembles considered previously if the expected degrees
are all sufficiently smaller than $\sqrt{2E}$, however it retains a lack
of intrinsic degree correlations even if this condition is not
satisfied, since the expected number of edges between nodes is always a
product of the fugacities. More specifically, the expected degree
$\avg{k}_\text{nn}(k)$ of the neighbors of nodes of degree $k$ is given
by
\begin{align}
  \avg{k}_\text{nn}(k) &= \sum_{\A}P(\A)\frac{\sum_i\delta_{k_i,k}\sum_jA_{ji}k_j/k_i}{\sum_i\delta_{k_i,k}}\\
  &= \sum_i \theta_i^2 = \frac{\sum_i \hat k_i^2}{2E},
\end{align}
which is a constant independent of $k$.

The Poisson model has been proposed originally by Norros and
Reittu~\cite{norros_conditionally_2006}, but not as a maximum-entropy
ensemble for multigraphs possessing distinguishable multiedges, as we do
here. At first, this might seem like a construct designed primarily for
mathematical convenience, rather than a principled proposition, as there
is no inherent property of a multigraph that allows us to tell
multiedges apart. However, there are situations where the notion of
multiedge distinguishability does arise naturally. For example, there
may be different roads between the same two
cities~\cite{barthelemy_spatial_2011}, both of which are identifiable
due to their spacial location. In proximity
networks~\cite{cattuto_dynamics_2010}, multi-edges correspond to events
that are localized in space and time, and hence are distinguishable. We
consider in Appendix~\ref{app:equivalence} different kinds of intuitive
random graph models that possess this property and show how they are
exactly equivalent to the Poisson model. Nevertheless it generates
multigraphs, where in many realistic settings we are interested in
simple graphs, with at most one edge between two nodes.

A common approach is simply to ignore the discrepancy, and employ the
Poisson model even when modeling simple
graphs~\cite{karrer_stochastic_2011}, arguing that the difference is
negligible for sparse graphs, a point which we will examine in more
detail in Sec.~\ref{sec:community}. For now, we simply anticipate that
in order for this approximation to be valid, the graphs need to be
\emph{uniformly} sparse. In the following, we consider the erased
Poisson model, which provides a better alternative to model simple
graphs with heterogeneous sparsity.

\section{The erased Poisson model for simple graphs}\label{sec:erased}

An alternative to the maximum-entropy model for simple graphs is the
``erased'' Poisson model, where a multigraph $\A$ is generated from the
Poisson model $P(\A|\bm\theta)$ and a simple graph $\G(\A)$ is obtained
from it by simply ignoring (``erasing'') multiedge multiplicities and
removing self-loops~\cite{vanden_esker_universality_2008,
bhamidi_scaling_2010}, i.e.
\begin{equation}\label{eq:erasing}
  G_{ij}(A_{ij}) =
  \begin{cases}
    1 & \text{ if } A_{ij} > 0 \text{ and } i\ne j,\\
    0 & \text{ otherwise.}
  \end{cases}
\end{equation}
The resulting simple graph $\G$ is generated with probability
\begin{equation}\label{eq:erased_poisson}
  P(\G|\bm\theta) = \prod_{i<j}\left(1-\e^{-\theta_i\theta_j}\right)^{G_{ij}}\e^{-\theta_i\theta_j(1-G_{ij})},
\end{equation}
and we can impose a desired expected degree sequence by solving the
system of equations
\begin{equation}\label{eq:poisson_constraints}
  \sum_{j\ne i}1-\e^{-\theta_i\theta_j} = \hat{k}_i.
\end{equation}
These equations also do not admit a general closed-form solution for
$\bm\theta$. We therefore may ask if this model is any more practical
than the maximum-entropy variant of Eq.~\ref{eq:maxent_simple}. As it
turns out, it is, but before we see how, let us compare the properties
of this model with the other ones we have seen so far.

\subsection{Degree correlations}\label{sec:correlations}

\begin{figure}
  \includegraphics[width=\columnwidth]{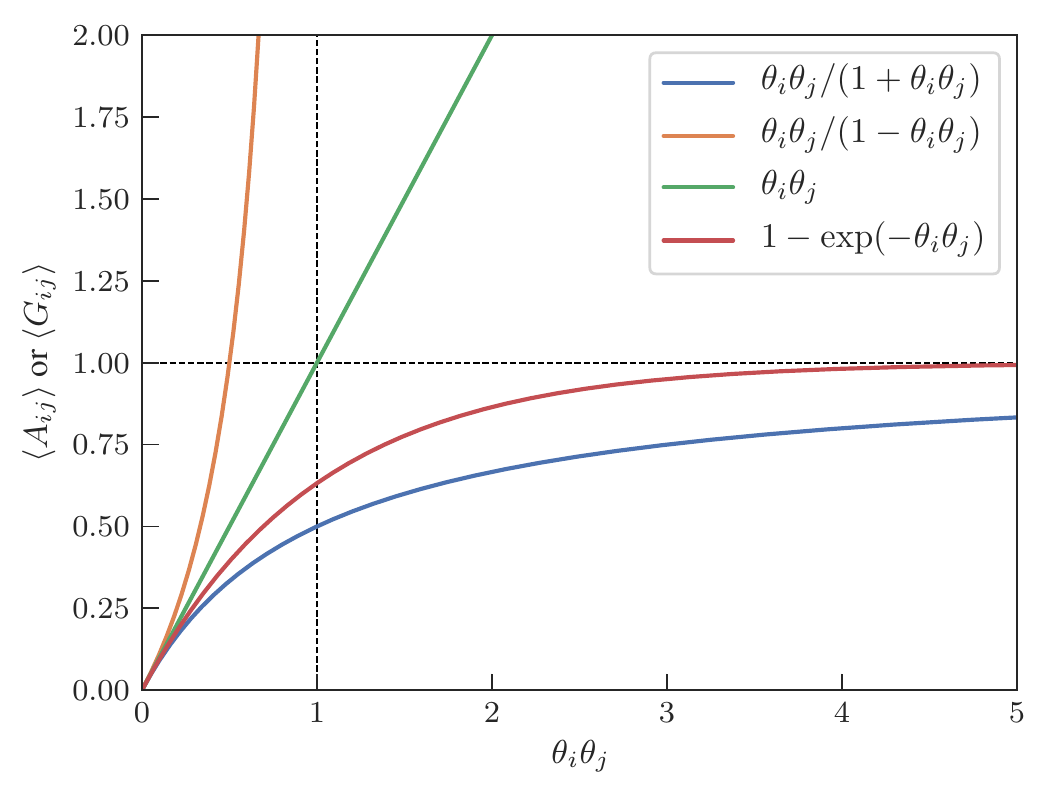} \caption{Average
  number of edges between nodes as a function of the product of their
  fugacities, for the different ensembles, as shown in the
  legend.\label{fig:placement}}
\end{figure}

With the exception of the Poisson model, all other models considered
yield samples with some form of degree-degree correlations. The
difference between the ensembles can seen by the inspecting the average
value of the adjacency matrix entries as a function of the product of
the fugacities in each case, i.e.
\begin{align}
  \avg{A_{ij}} &= \frac{\theta_i\theta_j}{1+\theta_i\theta_j}, & \text {(Max-Ent simple graph)}\nonumber\\
              &= \frac{\theta_i\theta_j}{1-\theta_i\theta_j}, & \text {(Max-Ent multigraph)}\nonumber\\
              &= \theta_i\theta_j, & \text {(Poisson multigraph)}\nonumber\\
\avg{G_{ij}}   &= 1-\e^{-\theta_i\theta_j}. & \text {(Erased Poisson simple graph)}\nonumber
\end{align}
These functions are illustrated in Fig.~\ref{fig:placement}.  For
$\theta_i\theta_j \ll 1$ all functions approach the same uncorrelated
placement of edges as in the Poisson model with $\avg{A_{ij}} \approx
\theta_i\theta_j$. For larger values the simple graph models show a
saturation of the edge placement probability, which results in a
disassortative degree correlation, as it prevents an excess of connections
to nodes with larger fugacities. The maximum-entropy multigraph model,
on the other hand, shows a divergence of the number of edges placed as
$\theta_i\theta_j\to 1$, which results in an assortative degree
correlation, due to the nonlinear accumulation of multiedges between
nodes with high fugacity.\footnote{This multiedge concentration is
reminiscent of the Bose-Einstein condensation phenomenon in quantum
physics, where the number of particles in the ground state of a Bose gas
diverges in a similar way. Indeed, Eq.~\ref{eq:avga_simple} for simple
graphs and Eq.~\ref{eq:avga_multi} for multigraphs follow the
Fermi-Dirac and Bose-Einstein statistics, respectively. Following this
analogy, the uniformly sparse graph regime where both ensembles agree
would correspond to the classical Maxwell-Boltzmann statistics, valid
for low densities or high temperatures. The Poisson multigraph model can
be interpreted as an extension of this classical limit to arbitrary
densities. We note however that, in network science, Bose-Einstein
condensation is more commonly associated with a different phenomenon of
growing networks~\cite{bianconi_bose-einstein_2001}.} (We note that it
is sometimes implied in the literature that multigraph ensembles are
uncorrelated. This is only true for models where multiedges are
distinguishable, such as the Poisson and configuration models, not
otherwise.)
\begin{figure}
  \begin{tabular}{cc}
    \begin{overpic}[width=.5\columnwidth]{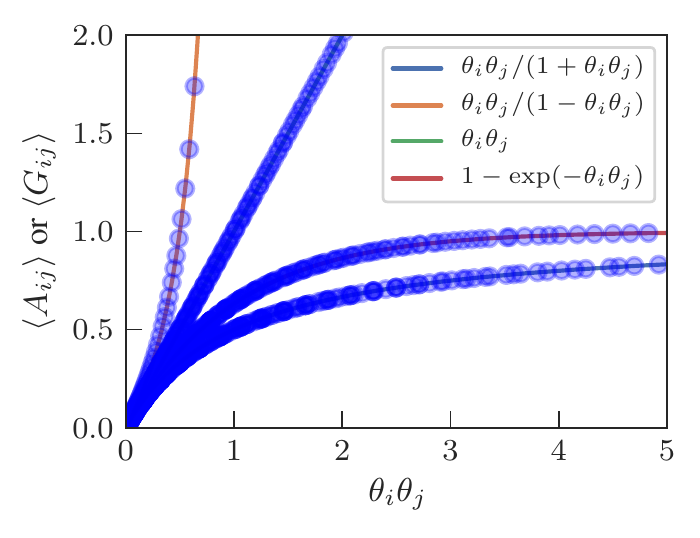}
      \put(0,0){(a)}
    \end{overpic}
    &
    \begin{overpic}[width=.5\columnwidth]{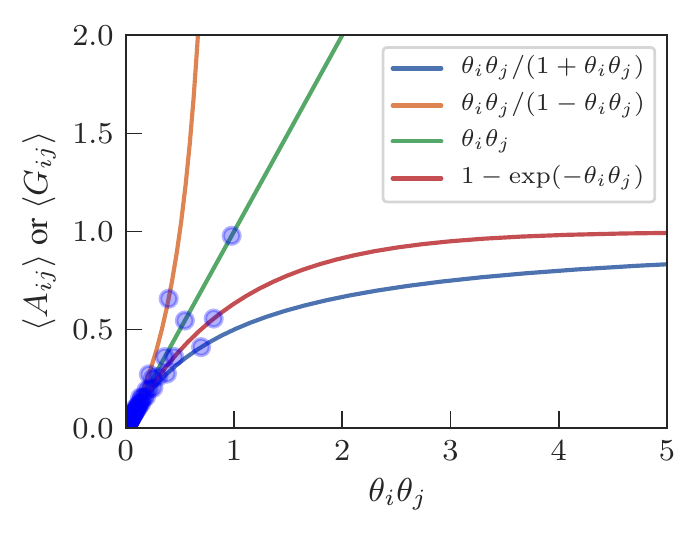}
      \put(0,0){(b)}
    \end{overpic}
  \end{tabular}

  \caption{The same as Fig.~\ref{fig:placement}, but showing the values
  for the edges of samples of each model with $N=10^6$ nodes, and with
  the same set of imposed degrees sampled from the Zipf distribution of
  Eq.~\ref{eq:zipf}, using
  (a) $\alpha=2.2$ and (b) $\alpha=2.4$.\label{fig:saturation}}
\end{figure}

For the Poisson multigraph model, the purely linear dependence on the
fugacities implies a total lack of correlations between degrees at the
endpoints of each edge, as we have already seen. The situation changes
when multigraphs are erased, where we observe a similar, although not
identical, saturation in the edge placement probability, which also
results in a disassortative degree-degree correlation among neighbors.

The difference between the ensembles can be further illustrated by
choosing integer-valued imposed degrees that are independently sampled
from a Zipf distribution
\begin{equation}\label{eq:zipf}
  P(k|\alpha) = \frac{k^{-\alpha}}{\zeta(\alpha)},
\end{equation}
with $\zeta(\alpha)$ being the Riemann zeta function. For values of
$\alpha\in[2,3]$ the variance of this distribution diverges, while the
mean remains finite, therefore serving as a simple model of globally
sparse but locally dense networks. In Fig.~\ref{fig:saturation} we show
how the edges present in one sample of each model are distributed along
the curves of Fig.~\ref{fig:placement}: A smaller value of $\alpha$
creates broader and denser networks, for which the discrepancy between
all models is very large. Although the mean degree of the generated
networks in the case $\alpha=2.2$ is only around $3.75$, even on a
network of $N=10^6$ nodes the probability of observing an edge between
two nodes approaches one for a significant number of pairs.  As the
exponent $\alpha$ increases, the network becomes more homogeneously
sparse, and the fugacities and corresponding edge probabilities become
more similar across ensembles.

\begin{figure}
  \includegraphics[width=\columnwidth]{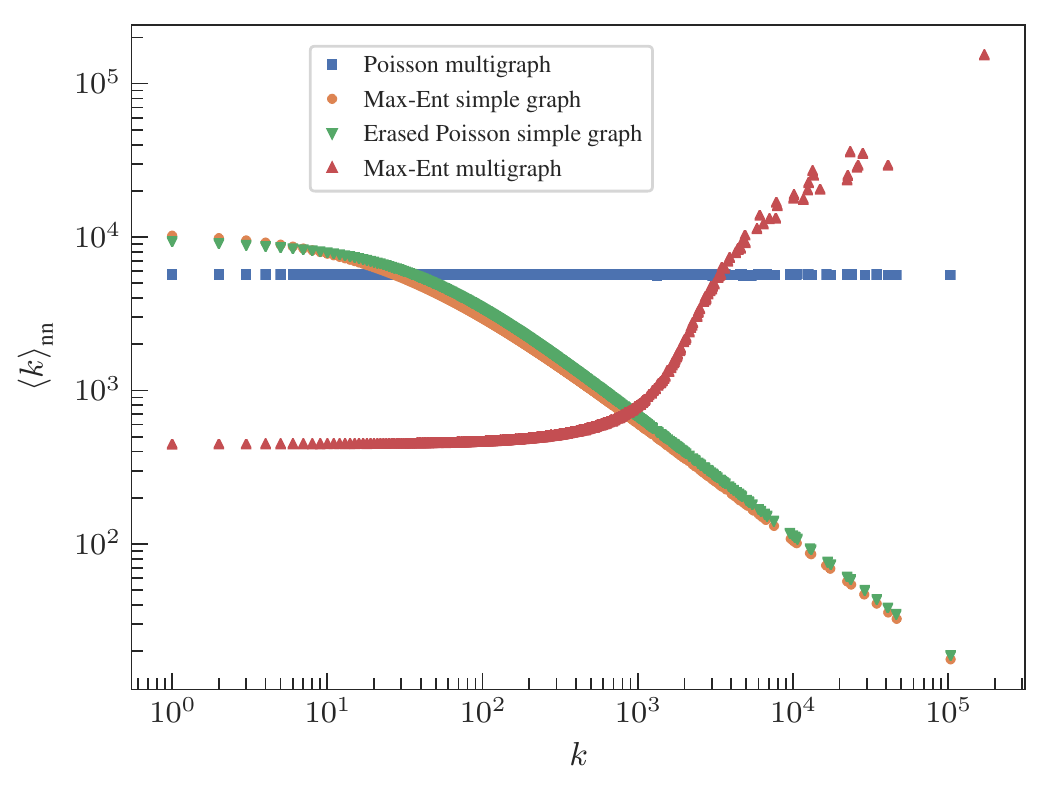} \caption{Mean
  degree of a neighbor of a node of degree $k$, as a function of $k$,
  $\avg{k}_{nn}(k)$, for the different ensembles considered, with
  $N=10^6$ and the same set of imposed degrees sampled from
  Eq.~\ref{eq:zipf} with $\alpha=2.2$. (The error bars on the
  $\avg{k}_{nn}$ values are smaller than the
  symbols used.) \label{fig:deg-corr}}
\end{figure}

The induced degree correlations among neighbors in each ensemble are
shown in Fig.~\ref{fig:deg-corr}, in each case for the same set of
imposed degrees sampled from Eq.~\ref{eq:zipf}. Some aspects of the
degree correlations of the erased Poisson model were considered
rigorously in Refs.~\cite{yao_average_2017,stegehuis_degree_2019}, and
in the case of the maximum-entropy simple graph model in
Ref.~\cite{park_origin_2003}. One important point to notice is that,
although the erased Poisson and maximum-entropy models for simple graphs
are not identical, they generate a very similar disassortative trend,
indicating a comparable explanatory power for this kind of effect.

\begin{figure}
  \includegraphics[width=\columnwidth]{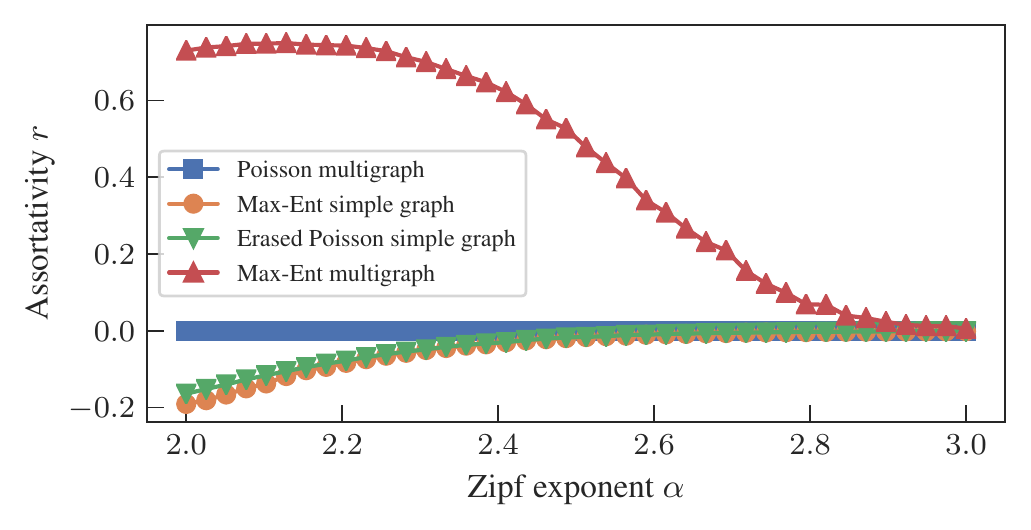} \caption{Degree
  assortativity $r$ as a function of the Zipf exponent $\alpha$, for
  networks with $N=10^6$ nodes sampled from the models indicated in the legend.
  \label{fig:assort}}
\end{figure}

A further comparison is seen in Fig.~\ref{fig:assort} where the
degree assortativity coefficient~\cite{newman_mixing_2003} $r \in
[-1,1]$ is shown as a function of the Zipf exponent $\alpha$, defined
as
\begin{equation}
  r = \frac{\sum_{kk'}kk'(m_{kk'}-q_kq_{k'})}{\sigma_k^2},
\end{equation}
with $m_{kk'} = \sum_{ij}A_{ij}\delta_{k_i,k}\delta_{k_j,k'}/2E$ being
the fraction of edges between nodes of degrees $k$ and $k'$,
$q_k=\sum_{k'}m_{kk'}$ is distribution of degrees at the endpoints of
edges, and $\sigma_k^2$ is its variance. We see that the
same pattern persists for the entire range of $\alpha \in [2,3]$, with
the assortativity values being very similar between the erased Poisson
and maximum-entropy simple graph models.

The similarity between the erased Poisson and the maximum-entropy simple
graph model makes the former attractive as an alternative, considering its
flexibility, as we are about to explore in the next section.

\subsection{Reconstructing the erased Poisson model}\label{sec:reconstruction}

The ensembles considered previously translate different assumptions
about the data into probability distributions, conditioned on desired
constraints, and mediated by the maximum-entropy ansatz. In principle we
should choose the set of assumptions that most closely matches the data
in question.

Although we have argued that the Poisson model is well motivated in
situations where multiedges can be distinguished, for simple graphs in
particular there is a practical advantage to using the erased Poisson
model, regardless if it is truly the most adequate mechanistic
explanation or null model. Namely, it allows us to disentangle the edge
placement from the inherent degree correlations, since the latter is
only caused by the erasing of multiedges, and is absent from the
original multigraph. We can do the same even if we only have the final
simple graph at our disposal, by attempting to reconstructing the
multigraph that generated it. If the latent multigraph does not exhibit
degree correlations, we can conclude those were caused by the erasure
procedure --- which simply reflects the inherent constraints of having
to generate a simple graph, and not some extrinsic propensity of
high-degree nodes to connect to low-degree ones.

In the following we describe a principled and efficient method to
perform such a reconstruction. We approach the task in a Bayesian way,
by considering the posterior distribution of multigraphs $\A$ and
fugacities $\bm\theta$ conditioned on an observed simple graph $\G$,
\begin{align}
  P(\A,\bm\theta|\G) = \frac{P(\G|\A)P(\A|\bm{\theta})P(\bm{\theta})}{P(\G)},
\end{align}
with $P(\G|\A)$ given by
\begin{align}\label{eq:erasing_indicator}
  P(G_{ij}|A_{ij}) &=
  \begin{cases}
    1, & \text{ if } G_{ij} = G_{ij}(A_{ij}),\\
    0, & \text{ otherwise},
  \end{cases}
\end{align}
with $G_{ij}(A_{ij})$ given by Eq.~\ref{eq:erasing} and with
$P(\A|\bm\theta)$ being the Poisson multigraph model of
Eq.~\ref{eq:poisson}. The distribution $P(\bm\theta)$ is our prior for
the fugacities, which for the moment we will assume to be constant
$P(\bm\theta)\propto 1$, meaning we are fully agnostic about what kind
of model generated the data (we will revisit this assumption in
Sec.~\ref{sec:community}). Finally, we have the so-called evidence
\begin{align}
  P(\G) &= \sum_{\A}\int P(\G|\A)P(\A|\bm\theta)P(\bm\theta)\,\dd\bm\theta,
\end{align}
which is an unimportant constant for our present purpose. With the
posterior distribution $P(\A,\bm\theta|\G)$ in place, we can proceed in
a variety of ways, for example by sampling from it using MCMC. But
instead, we will proceed in a more efficient manner, by considering
first the most likely fugacities, when averaged over all possible
multigraphs $\A$, i.e.
\begin{equation}
  \hat{\bm\theta} = \underset{\bm\theta}{\operatorname{argmax}}\,\sum_{\A}P(\A,\bm\theta|\G) = \underset{\bm\theta}{\operatorname{argmax}}\,P(\G|\bm\theta),
\end{equation}
where $P(\G|\bm\theta)$ is the erased Poisson likelihood of
Eq.~\ref{eq:erased_poisson}. Noting that taking the logarithm of the
likelihood does not alter the position of its maximum, and substituting leads to
\begin{equation}
  \hat{\bm\theta} = \underset{\bm\theta}{\operatorname{argmax}}\,\sum_{i< j}G_{ij}\ln\left(1-\e^{-\theta_i\theta_j}\right) - (1-G_{ij})\theta_i\theta_j.
\end{equation}
Taking the derivatives of the right hand side with respect to
$\bm\theta$ and setting them to zero yields a system of nonlinear
implicit equations that does not admit an obvious solution in closed
form in the general case. Fortunately, we can obtain a simple algorithm
for solving it, by slightly augmenting our problem, and obtaining at the
same time a conditional posterior distribution over multigraphs
$P(\A|\G,\hat{\bm{\theta}})$. We do so by employing Jensen's inequality
on $P(\G|\bm\theta) = \sum_{\A}P(\G|\A)P(\A|\bm\theta)$ in the form
\begin{align}
\ln \sum_{\A}P(\G|\A)P(\A|\bm\theta) \ge \sum_{\A}q(\A) \ln \frac{P(\G|\A)P(\A|\bm\theta)}{q(\A)},
\end{align}
where the equality is achieved by setting
\begin{align}
  q(\A) &= \frac{P(\G|\A)P(\A|\bm\theta)}{\sum_{\A'}P(\G|\A')P(\A'|\bm\theta)} = P(\A|\G,\bm\theta),\\
       &= \prod_{i\le j}P(A_{ij}|G_{ij},\bm\theta),
\end{align}
which is precisely the posterior distribution of multigraphs conditioned
on a particular choice of fugacities, whose entries can be directly computed as
\begin{equation}
  P(A_{ij}|G_{ij},\bm\theta) =
  \begin{cases}
    \frac{\theta_i\theta_j^{A_{ij}}\e^{-\theta_i\theta_j}}{A_{ij}!}\frac{1-\delta_{A_{ij},0}}{1-\e^{-\theta_i\theta_j}}& \text{ if } G_{ij} = 1,\\
    \frac{(\theta_i^2/2)^{A_{ii}/2}\e^{-\theta_{ii}^2/2}}{(A_{ii}/2)!} & \text{ if } i = j,\\
    0 & \text{ otherwise.}
  \end{cases}
\end{equation}
It will be useful to summarize this posterior distribution via its mean
value for each node pair, given by
\begin{equation}
  w_{ij} \equiv \avg{A_{ij}} =
  \begin{cases}
    \frac{\theta_i\theta_j}{1-\e^{-\theta_i\theta_j}}& \text{ if } G_{ij} = 1,\\
    \theta_i^2 & \text{ if } i = j,\\
    0 & \text{ otherwise.}
  \end{cases}
\end{equation}
With this at hand, we then return to the maximization to obtain
\begin{align}
  \hat{\bm\theta}
   &= \underset{\bm\theta}{\operatorname{argmax}}\; \sum_{\A} q(\A) \ln \frac{P(\G|\A)P(\A|\bm\theta)}{q(\A)}\\
   &= \underset{\bm\theta}{\operatorname{argmax}}\; \sum_{i\le j}\sum_{A_{ij}=0}^{\infty}  q(A_{ij}) \ln P(A_{ij}|\bm\theta)\\
   &= \underset{\bm\theta}{\operatorname{argmax}}\; \frac{1}{2}\sum_{ij}w_{ij}\ln\theta_i\theta_j - \theta_i\theta_j.
\end{align}
The last equation can be solved easily, which yields
\begin{equation}
  \hat\theta_i = \frac{d_i}{\sqrt{\sum_jd_j}},
\end{equation}
where
\begin{equation}
  d_i = \sum_jw_{ji}
\end{equation}
is the expected degree of node $i$ in the multigraph $\A$, averaged over
$P(\A|\G,\hat{\bm{\theta}})$. Since we are interested in the
self-consistent values of $\bm w$ conditioned on $\hat{\bm{\theta}}$,
this leads us to the following expectation-maximization (EM) algorithm,
which starts with some arbitrary choice of $\bm\theta$, and alternates
between the following steps
\begin{enumerate}
\item In the ``expectation'' step we obtain the marginal mean multiedge
      multiplicities via:
\begin{equation}
  w_{ij} =
  \begin{cases}
    \frac{\theta_i\theta_j}{1-\e^{-\theta_i\theta_j}} & \text{ if } G_{ij} = 1,\\
    \theta_i^2 & \text{ if } i = j,\\
    0 & \text{ otherwise.}
  \end{cases}
\end{equation}
\item In the ``maximization'' step we use the current values of $\bm w$ to
      update the values of $\bm\theta$:
      \begin{equation}
        \theta_i = \frac{d_i}{\sqrt{\sum_jd_j}}, \quad\text{ with } d_i = \sum_jw_{ji}. 
      \end{equation}
\end{enumerate}
Upon convergence, the above EM algorithm is guaranteed to find only a
local optimum of the maximization problem, therefore we may need to run
it multiple times with different initial choices of
$\bm\theta$. However, in all our experiments we found that the algorithm
tends to find the same solution from any initial starting point, even
when this happens to be the correct solution (in artificially generated
examples where this is known), giving strong evidence that the global
optimum is usually found. This algorithm is efficient, since we need
only to keep track of the values for $\bm w$ for the observed edges in
$\bm G$, in addition to each self-loop. Therefore the E-step can be done
in $O(E+N)$ time, and the M-step in $O(N)$ time, resulting in an overall
$O(E+N)$ computational complexity. The algorithm can also be run in
parallel easily. The number of EM iterations required for convergence
depends on the data and initial conditions, but we have successfully run
it on networks with up to $10^8$ edges on a regular laptop computer. Our
C++ implementation of the above algorithm is available as part of the
\texttt{graph-tool} Python library~\cite{peixoto_graph-tool_2014}.

\begin{figure}
  \includegraphics[width=\columnwidth]{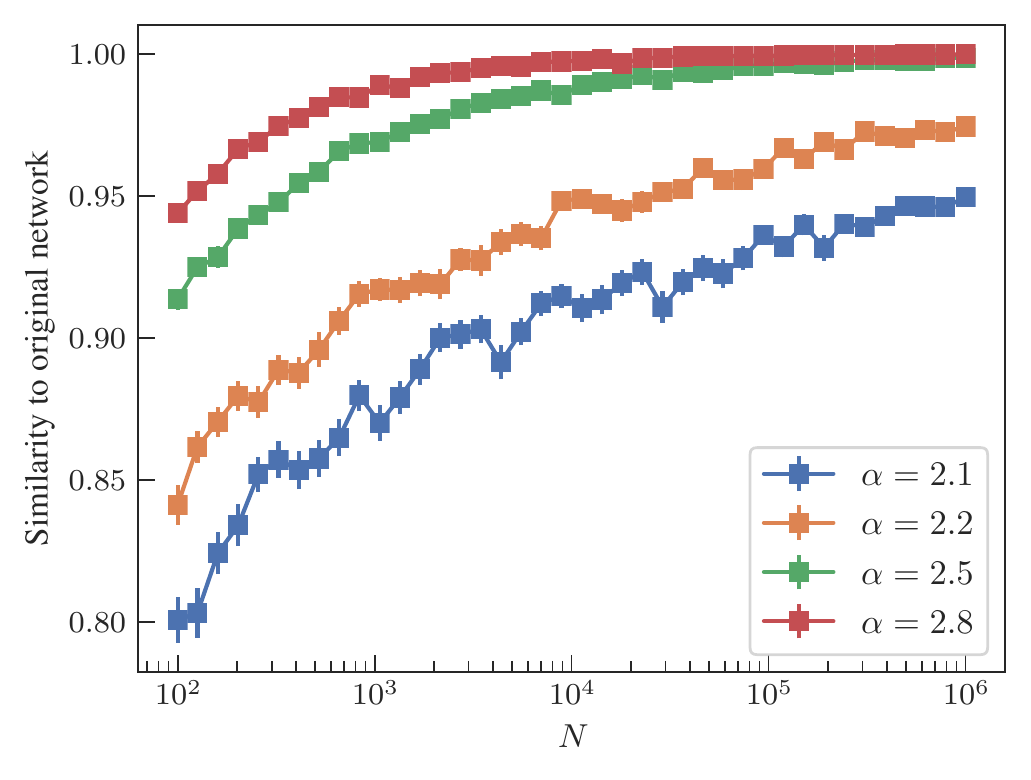}

  \caption{Poisson multigraph reconstruction accuracy as measured via
  the similarity of Eq.~\ref{eq:similarity} for simple graphs sampled
  from the erased Poisson model with imposed degrees sampled from a Zipf
  distribution with exponent $\alpha$, as a function of the number of
  nodes $N$. Each point was averaged over 100
  realizations.\label{fig:recovery}}
\end{figure}

\begin{figure}
  \includegraphics[width=\columnwidth]{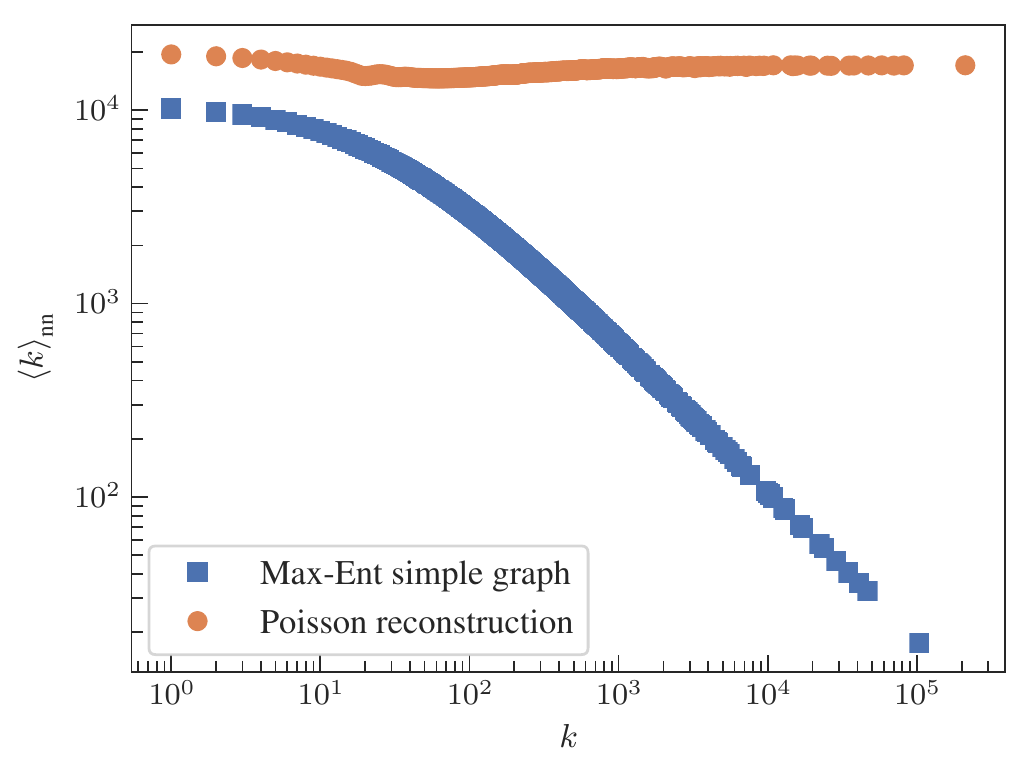}
  \caption{Mean degree of a neighbor of a node of degree $k$, as a
  function of $k$, for a network with $N=10^6$ nodes sampled from the
  maximum-entropy ensemble with imposed degrees sampled from
  Eq.~\ref{eq:zipf} with $\alpha=2.2$, and its inferred Poisson
  multigraph, using the algorithm described in the
  text.\label{fig:rec-max-ent}}
\end{figure}

\begin{figure}
  \includegraphics[width=\columnwidth]{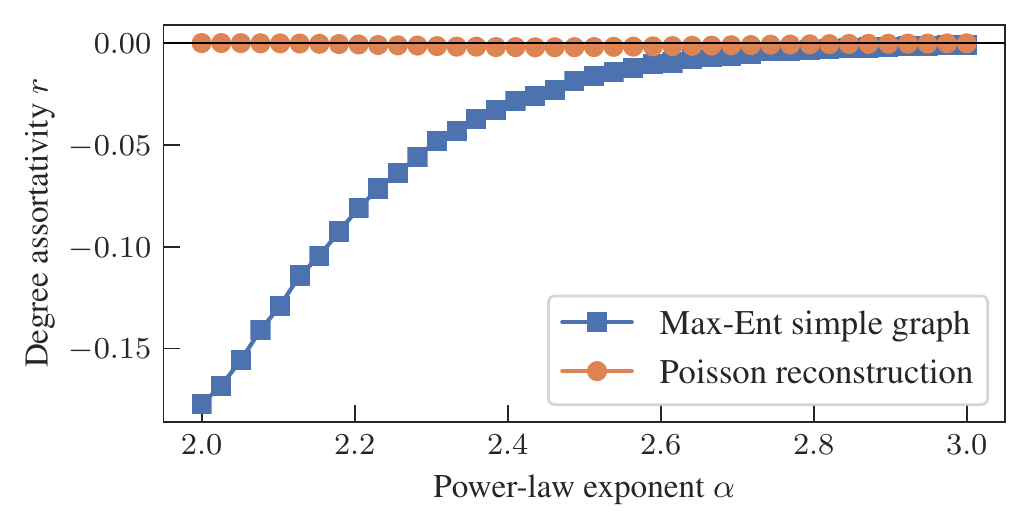}

  \caption{Degree assortativity as function of the Zipf exponent
  $\alpha$, for networks with $N=10^6$ nodes sampled from the maximum
  entropy ensemble with Zipf-distributed imposed degrees, and their
  corresponding inferred Poisson multigraphs.\label{fig:rec-max-ent-r}}
\end{figure}

In Fig.~\ref{fig:recovery} we show how the algorithm behaves in
recovering the underlying multigraph of artificial networks sampled from
the erased Poisson model, as measured via the Jaccard similarity
\begin{equation}\label{eq:similarity}
  s(\bm w, \A) = 1-\frac{\sum_{ij}|w_{ij}-A_{ij}|}{\sum_{ij}w_{ij}+A_{ij}}
\end{equation}
between the true and inferred multigraphs, with imposed degrees sampled
from a Zipf distribution. For smaller values of the exponent $\alpha$,
which causes the edge multiplicities to become larger, the recovery
becomes less accurate, but in all cases it approaches $s(\bm w,\A)\to 1$
as the number of nodes increases, indicating that full recovery is
possible asymptotically as the amount of available data increases.

Since this algorithm gives us a distribution over multigraphs, we can
use it to investigate whether the degree correlations of an observed
simple graphs exist as a necessary outcome of the existing degrees, or
if they should be attributed to something else. In the first scenario,
the degree-degree correlations would disappear in the inferred
multigraph, whereas they would persist in the second one. Interestingly,
this works reasonably well even when the observed graph was not sampled
from the erased Poisson model. We illustrate this with an example in
Fig.~\ref{fig:rec-max-ent}, where a simple graph was generated from a
maximum-entropy model with Zipf-distributed degrees, and we inferred
from it a corresponding Poisson multigraph. Even though the inferred
multigraph still shows a weak degree correlation between neighboring
nodes, since the erased Poisson model cannot fully account for the
structure of the maximum-entropy model, the overall disassortative trend
is completely absent. Fig.~\ref{fig:rec-max-ent-r} shows the degree
assortativity values for both original and reconstructed networks over a
range of $\alpha\in[2,3]$. Although the reconstructed networks still
show values $r < 0$, their deviation from zero is barely noticeable in
the figure. Like we had seen previously in Fig.~\ref{fig:deg-corr},
these results further show that the erased Poisson model generates
mixing patterns that, although not identical, are sufficiently similar
to the maximum-entropy simple graph model, allowing us to correctly
conclude that the resulting degree correlations arise directly from the
imposed degrees.

\subsection{Empirical networks}
\begin{figure}
  \includegraphics[width=\columnwidth]{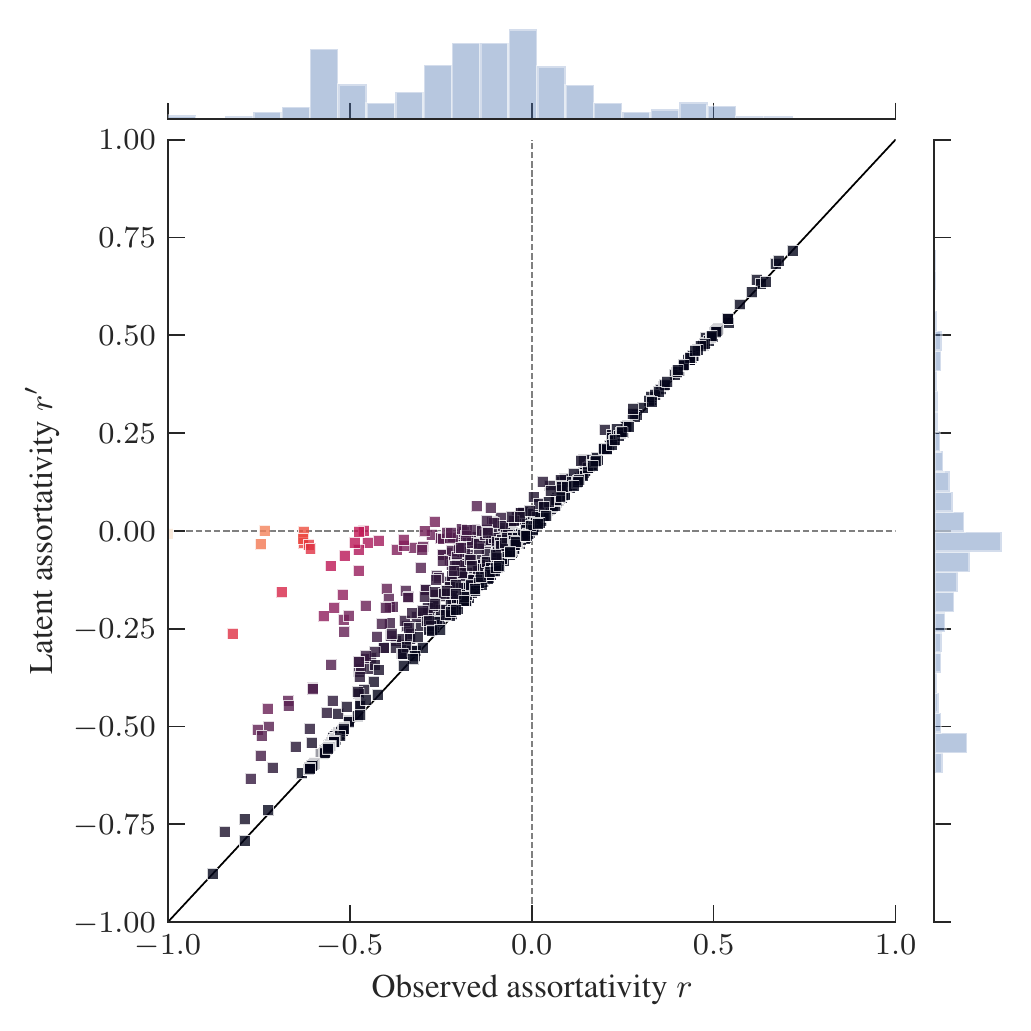}

  \caption{Degree assortativity for original simple graph $r$, and for
  the reconstructed multigraph $r'$ for 816 empirical networks, obtained
  from the CommunityFitNet~\cite{ghasemian_evaluating_2019} and
  Konect~\cite{kunegis_konect:_2013} databases.\label{fig:empirical}}
\end{figure}

\begin{figure}
  \includegraphics[width=\columnwidth]{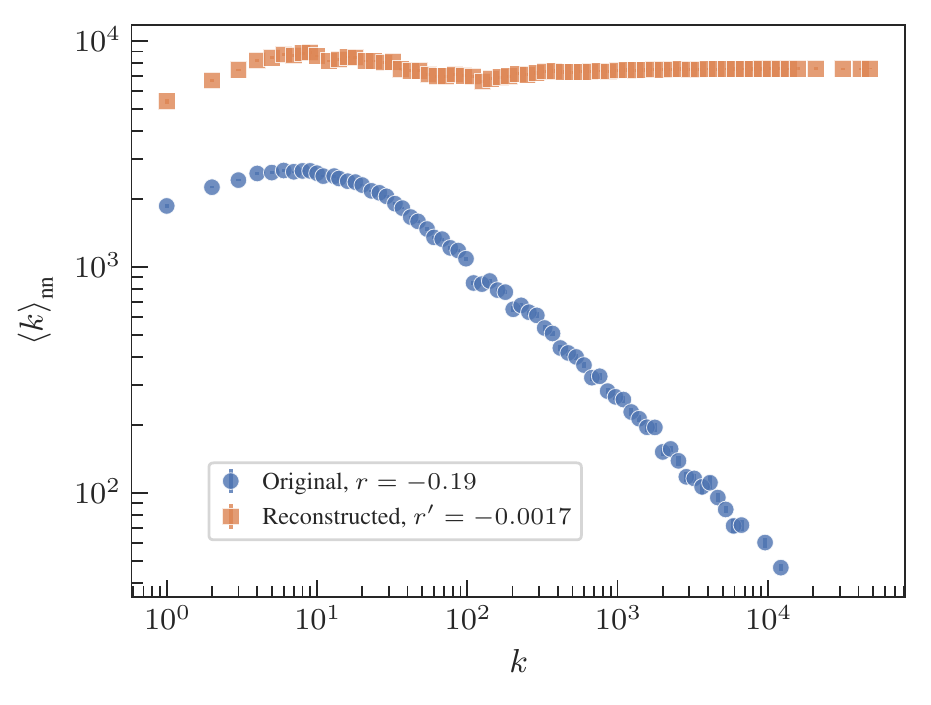}
  \caption{Mean degree of a neighbor of a node of degree $k$ as a
  function of $k$, for the internet at the autonomous systems level,
  both for the original network and reconstructed multigraph, as shown
  in the legend, which includes the degree assortativity coefficient of
  each case.\label{fig:caida}}
\end{figure}

\begin{figure}
  \begin{tabular}{cc}
    \begin{overpic}[width=.5\columnwidth]{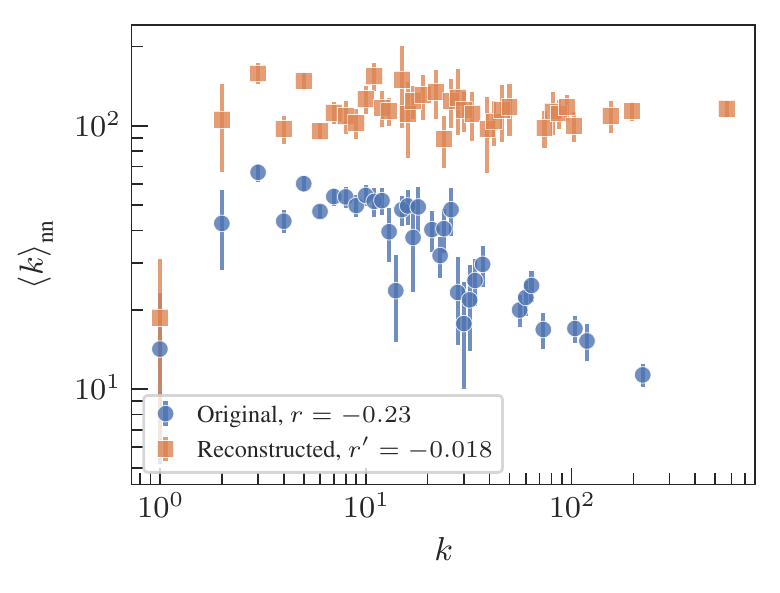}
      \put(0,0){(a)}
    \end{overpic}&
    \begin{overpic}[width=.5\columnwidth]{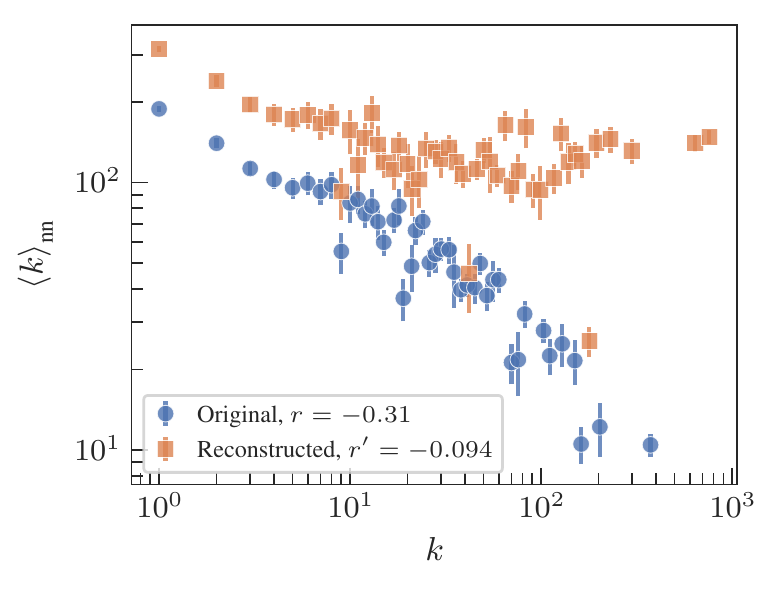}
      \put(0,0){(b)}
    \end{overpic}\\
    \begin{overpic}[width=.5\columnwidth]{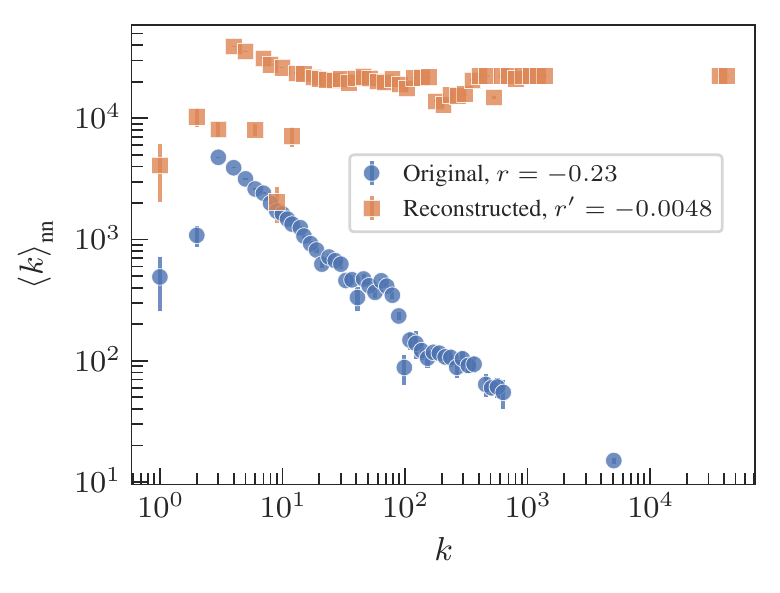}
      \put(0,0){(c)}
    \end{overpic}&
    \begin{overpic}[width=.5\columnwidth]{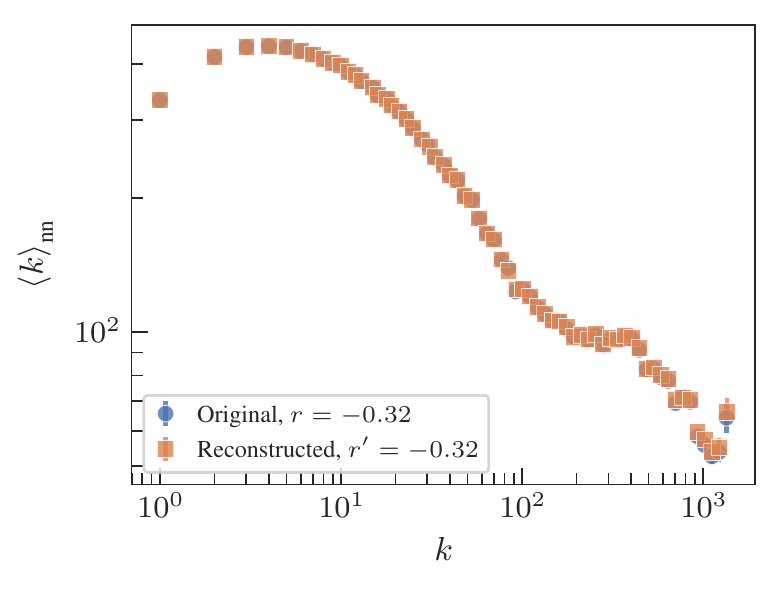}
      \put(0,0){(d)}
    \end{overpic}
  \end{tabular}
  \caption{Mean degree of a neighbor of node of degree $k$ as a
    function of $k$, for (a) the metabolic network of
    \emph{C. elegans}~\cite{duch_community_2005}, (b) the network of
    leaked emails of the Democratic National Committee, (c) the class
    dependency graph of a large software
    project~\cite{subelj_software_2012}, and (d) the online social
    network Flixter~\cite{zafarani_social_2009}, both for the original
    network and reconstructed multigraph, as shown in the legend, which
    includes the degree assortativity coefficient of each
    case.\label{fig:several}}
\end{figure}

We can use the erased Poisson model to decouple degree assortativity
from the degree constraints by inferring the underlying multigraph where
these properties are not tied to each other. In Fig.~\ref{fig:empirical}
we show the results of applying our algorithm to 816 networks across
different domains, obtained from the
CommunityFitNet~\cite{ghasemian_evaluating_2019} and
Konect~\cite{kunegis_konect:_2013} databases, and comparing the
assortativity coefficient computed for the original network and
reconstructed multigraph. For assortative mixing patterns with $r>0$ we
do not observe any significant difference, as this kind of mixing
pattern is unrelated to degree constraints. However, for disassortative
values $r<0$ we observe a variety of behaviors, where for many networks
the assortativity value is significantly increased in the reconstructed
multigraph, indicating that observed correlations can be largely
associated with the broadness of the degrees. A prime example of this is
the Internet at the autonomous systems
level~\cite{peixoto_hierarchical_2014}, which has been long considered
as a case where the observed disassortativity is a byproduct of the broad
degree sequence rather than an independent feature of the
network~\cite{maslov_detection_2004,park_origin_2003}. As we see in
Fig.~\ref{fig:caida}, we can recover this result clearly with our
reconstruction method, where the inferred multigraph completely lacks
the disassortativity pattern. Other examples of this phenomenon are show
in Fig.~\ref{fig:several} for the metabolic network of
\emph{C. elegans}~\cite{duch_community_2005}, the network of leaked
emails of the Democratic National Committee, and the class dependency
graph of a large software project~\cite{subelj_software_2012}. We also
show the reconstruction results for the now-extinct online social
network Flixter~\cite{zafarani_social_2009}, where users could share
their taste on films. This network displays a strong degree
disassortativity which persists completely in the reconstructed Poisson
multigraph, indicating that it does not in fact arise from the inherent
constraints of the existing degrees, and must therefore be due to some
other mechanism.

\section{Community detection}\label{sec:community}

Another important type of heterogeneous sparsity is community
structure~\cite{fortunato_community_2010}, which can be loosely defined
as the existence of groups of nodes with a high probability of
connection to themselves, or also to other groups. Models for networks
with this kind of structure can be obtained by forcing the number of
edges between groups to have specific values. More precisely, we assume
the nodes are divided into $B$ disjoint groups, with $b_i\in[1,B]$
denoting the group membership of node $i$. With this, and in addition to
the expected degree constraints, we have the expected edge counts
between groups given by
\begin{equation}
  \sum_{\A}P(\A)\sum_{ij}A_{ij}\delta_{b_i,r}\delta_{b_j,s} = m_{rs}.
\end{equation}
Performing the same calculation as before, i.e. maximizing the ensemble
entropy conditioned on the above constraints in addition to
Eq.~\ref{eq:const_k} we arrive at the model
\begin{equation}\label{eq:sbm_maxent}
  P(\A|\bm\lambda,\bm\theta,\bb) = \prod_{i<j} \frac{(\lambda_{b_ib_j}\theta_i\theta_j)^{A_{ij}}}{\lambda_{b_ib_j}\theta_i\theta_j+1},
\end{equation}
in the case of simple graphs, which contains another set of fugacities
$\bm\lambda$. The values of the fugacities $\bm\theta$ and $\bm\lambda$
are determined by solving the following set of equations
\begin{equation}
  \sum_{j\ne i}\frac{\lambda_{b_ib_j}\theta_i\theta_j}{\lambda_{b_ib_j}\theta_i\theta_j+1} = \hat k_i, \quad
  \sum_{ij}\frac{\lambda_{rs}\theta_i\theta_j\delta_{b_i,r}\delta_{b_j,s}}{\lambda_{rs}\theta_i\theta_j+1} = m_{rs},
\end{equation}
which once more cannot be solved in closed form.

Instead, if we consider multigraphs with distinguishable multiedges,
performing the same calculations as before, we arrive at the Poisson
version of the degree-corrected SBM (DC-SBM), originally proposed by
Karrer and Newman~\cite{karrer_stochastic_2011}
\begin{multline}
  P(\A|\bm\lambda,\bm\theta,\bb) = \prod_{i<j}\frac{(\lambda_{b_ib_j}\theta_i\theta_j)^{A_{ij}}\e^{-\lambda_{b_ib_j}\theta_i\theta_j}}{A_{ij}!}\times\\
  \prod_i\frac{(\lambda_{b_ib_i}\theta_i^2/2)^{A_{ij}/2}\e^{-\lambda_{b_ib_i}\theta_i^2/2}}{(A_{ij}/2)!}.
\end{multline}
As pointed out in Ref.~\cite{karrer_stochastic_2011}, this model is more
tractable, and we can obtain the fugacities directly as
\begin{equation}
  \theta_i = \frac{\hat k_i}{\sum_j\hat k_j\delta_{b_j,b_i}}, \quad \lambda_{rs} = m_{rs}.
\end{equation}

Rather than the fugacities, in this context we are primarily interested
in obtaining the partition $\bb$ given an observed network $\A$, hence
we focus on the posterior
\begin{equation}\label{eq:posterior}
  P(\bb|\A) = \frac{P(\A|\bb)P(\bb)}{P(\A)},
\end{equation}
with the marginal likelihood being integrated over the fugacities
\begin{equation}
  P(\A|\bb) = \int P(\A|\bm\lambda,\bm\theta,\bb) P(\bm\theta|\bb) P(\bm\lambda|\bb)\;\dd\bm\theta\dd\bm\lambda,
\end{equation}
If we use noninformative priors
\begin{align}
  P(\bm\theta|\bb) &= \prod_r(n_r-1)!\,\delta\left(\textstyle\sum_i\theta_i\delta_{b_i,r}-1\right),\\
  P(\bm\lambda|\bb,\bar\lambda) &= \prod_{r<s}\e^{-\lambda_{rs}/\bar\lambda}/\bar\lambda\prod_r\e^{-\lambda_{rs}/2\bar\lambda}/2\bar\lambda,
\end{align}
we can compute the integral for the Poisson model
as~\cite{peixoto_nonparametric_2017}
\begin{multline}\label{eq:marginal_likelihood}
  P(\A|\bb) = \frac{\bar\lambda^E}{(\bar\lambda+1)^{E+B(B+1)/2}}\times\frac{\prod_{r<s}e_{rs}!\prod_r e_{rr}!!}{\prod_{i<j}A_{ij}!\prod_i A_{ii}!!} \times\\
  \prod_r\frac{(n_r-1)!}{(e_r+n_r-1)!}\prod_ik_i!,
\end{multline}
where $e_{rs}=\sum_{ij}A_{ij}\delta_{b_i,r}\delta_{b_j,s}$, and
$e_r=\sum_se_{rs}$. Although there are good reasons not to use such
uninformative priors~\cite{peixoto_nonparametric_2017}, the above
calculation illustrates how the Poisson model allows us to perform
computations that would be very difficult with the maximum-entropy
model. Going further, and exploiting the equivalence with the
microcanonical configuration model as was shown in
Ref.~\cite{peixoto_nonparametric_2017}, it is possible to replace these
priors by nested sequences of priors and hyperpriors that enhance our
capacity to identify small groups in large networks, more adequately
describe broad degree sequences, and uncover hierarchical modular
structures.

Despite these advantages, the Poisson DC-SBM model inherits all the
shortcomings of the Poisson model we considered previously, when applied
to simple graph data. In order to alleviate these problems, we may
therefore also employ the erased Poisson model for community detection,
with a likelihood
\begin{equation}
  P(\G|\bm\lambda,\bm\theta,\bb) = \prod_{i<j}(1-\e^{-\lambda_{b_ib_j}\theta_i\theta_j})^{G_{ij}}\e^{-\lambda_{b_ib_j}\theta_i\theta_j(1-G_{ij})}.
\end{equation}
This likelihood, however, makes the direct computation of the marginal
likelihood intractable, as it is not easy to perform the integral over
$\bm\lambda$ and $\bm\theta$. Instead, we proceed in a different manner,
by considering the joint likelihood of the simple graph $\G$ and its
underlying multigraph $\A$,
\begin{equation}
  P(\G,\A|\bm\lambda,\bm\theta,\bb) = P(\G|\A)P(\A|\bm\lambda,\bm\theta,\bb),
\end{equation}
with $P(\G|\A)$ given by Eq.~\ref{eq:erasing_indicator}. In this manner,
we can easily write the joint posterior distribution over the node
partition and multigraph
\begin{equation}
  P(\A,\bb|\G) = \frac{P(\G|\A)P(\A|\bb)}{P(\G)},
\end{equation}
which involves the same marginal likelihood $P(\A|\bb)$ we computed
previously. With this posterior at hand, we can proceed by sampling both
the partition $\bb$ and the latent multigraph $\A$ via MCMC. We do so by
starting with some initial choice for $\A$ and $\bb$, and performing
moves of the partition according to a proposal probability
$P(\bb'|\bb)$, and accepting it with the
Metropolis-Hastings~\cite{metropolis_equation_1953,hastings_monte_1970}
probability
\begin{equation}
  \min\left(1, \frac{P(\A,\bb'|\G)P(\bb|\bb')}{P(\A,\bb|\G)P(\bb'|\bb)}\right),
\end{equation}
otherwise we reject the move.  Likewise, for a current value of $\A$ and
$\bb$ we also perform move proposals for the latent multigraph with
probability $P(\A'|\A)$ and accept it with the analogous criterion
\begin{equation}
  \min\left(1, \frac{P(\A',\bb|\G)P(\A|\A')}{P(\A,\bb|\G)P(\A'|\A)}\right).
\end{equation}
By alternating between the two kinds of moves, this algorithm will
sample asymptotically from the target distribution $P(\A,\bb|\G)$, as
long as the move proposals allow us to visit every possible $(\A,\bb)$
configuration with nonzero probability. Importantly, when computing the
ratios above, it is not necessary to compute the intractable
normalization constant $P(\G)$, as it is the same in the numerator and
denominator, and hence cancels out. For the partition move proposal
$P(\bb'|\bb)$, we use the targeted move proposals described in
Ref.~\cite{peixoto_nonparametric_2017}. For the multigraph move proposal
$P(\A'|\A)$ we choose an edge $(i,j)$ in $\G$ uniformly at random, and
change the corresponding value of $A_{ij}$ by summing or subtracting $1$
with equal probability, unless that change would make $A'_{ij}=0$, which
is forbidden since $G_{ij}=1$. This amounts to
\begin{equation}
  P(A_{ij}'| A_{ij}) =
  \begin{cases}
    1 &\text{ if } A'_{ij} = 2 \text{ and } A_{ij} = 1, \\
    1/2 &\text{ if } A'_{ij} = A_{ij} \pm 1 \text{ and } A_{ij} > 1, \\
    0 &\text{ otherwise.}
  \end{cases}
\end{equation}
For the DC-SBM model above, a move $\bb\to\bb'$ that changes the group
membership of a single node can be done in time $O(k_i)$ where $k_i$ is
the degree of that node in $\G$, independently of how many groups exist
in total~\cite{peixoto_nonparametric_2017}. The move $A_{ij}\to
A_{ij}\pm 1$ can be done in constant time $O(1)$, as it involves the
change of at most a single value of $e_{rs}$, $e_r$ and $k_i$ in the
likelihood of Eq.~\ref{eq:marginal_likelihood} for each endpoint of the
edge (which remains true when the more advanced priors in
Ref.~\cite{peixoto_nonparametric_2017} are used instead). Therefore a
full sweep of move proposals for each node and edge in $\G$ can be done
in linear time $O(N + E)$, which is the best one can hope for this
problem, and enables the use of this algorithm for networks with
millions of nodes and edges. A reference C++ implementation of the above
algorithm is available as part of the \texttt{graph-tool} Python
library~\cite{peixoto_graph-tool_2014}.

We emphasize that sampling from the joint posterior $P(\A,\bb|\G)$ gives
us direct access to the marginal posterior over partitions as well,
\begin{equation}
  P(\bb|\G) = \sum_{\A}P(\A,\bb|\G),
\end{equation}
which can be obtained with the MCMC algorithm above simply by sampling
from the joint distribution, and ignoring the inferred multigraph
$\A$. So, if we are interested only in the community detection problem,
we are well served by this approach. However, obtaining the latent
multigraph $\A$ also has its uses, as we had seen before, for instance
in disentangling degree mixing from inherent degree constraints. We can
therefore also extract the marginal distribution over edge
multiplicities in an analogous way
\begin{equation}
  P(\A|\G) = \sum_{\bb}P(\A,\bb|\G).
\end{equation}
It is often more convenient to compute the marginal multiplicity
distribution over each edge
\begin{equation}
  \pi_{ij}(m) = \sum_{\A,\bb}\delta_{A_{ij},m}P(\A,\bb|\G),
\end{equation}
or more simply just its mean value
\begin{equation}
  w_{ij} = \sum_{m=0}^{\infty}m\pi_{ij}(m).
\end{equation}

\begin{figure*}
  \begin{tabular}{cc}
    \begin{overpic}[width=\columnwidth, trim=0 2cm 0 2cm, clip]{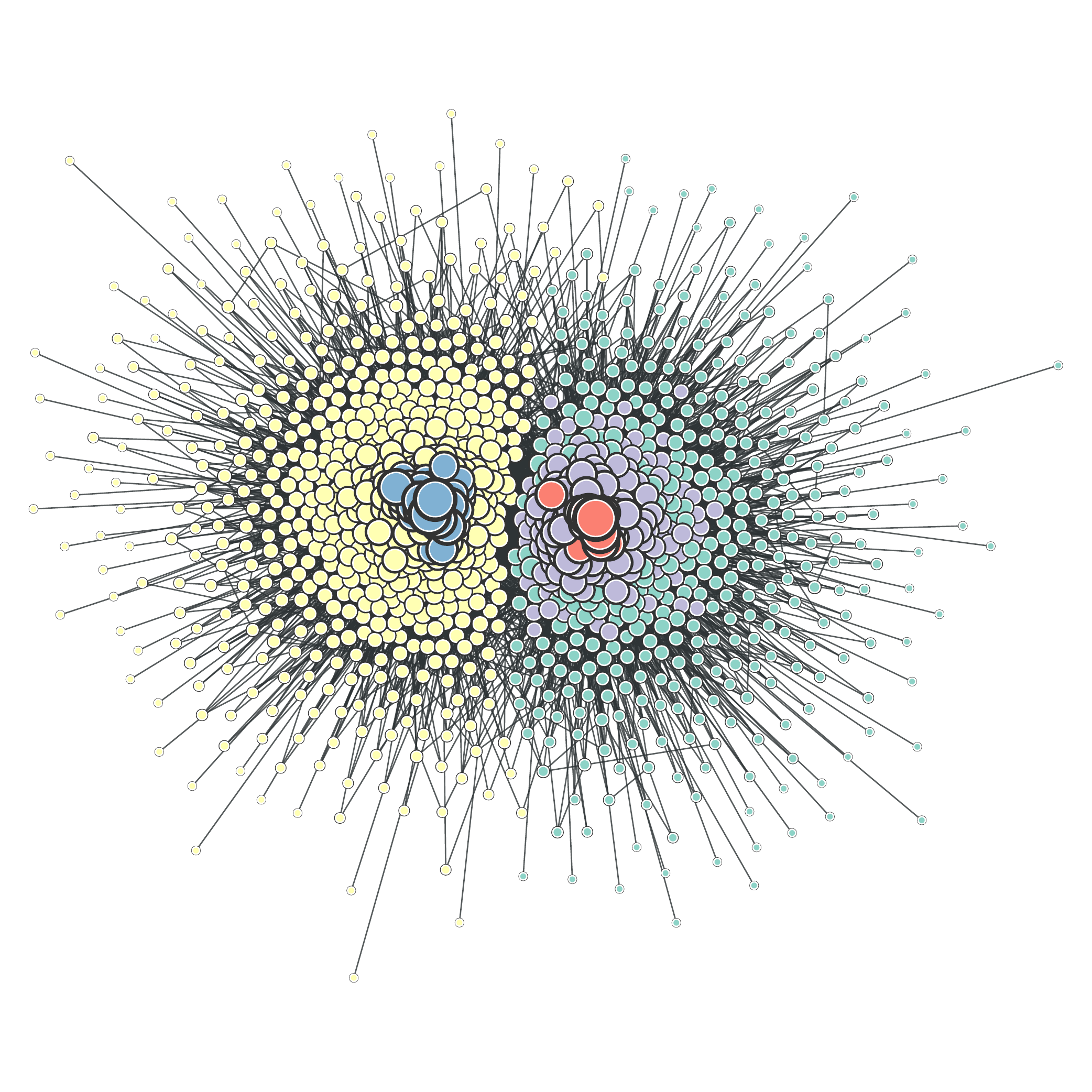}
      \put(0,0){(a)}
    \end{overpic} &
    \begin{overpic}[width=\columnwidth, trim=0 2cm 0 2cm, clip]{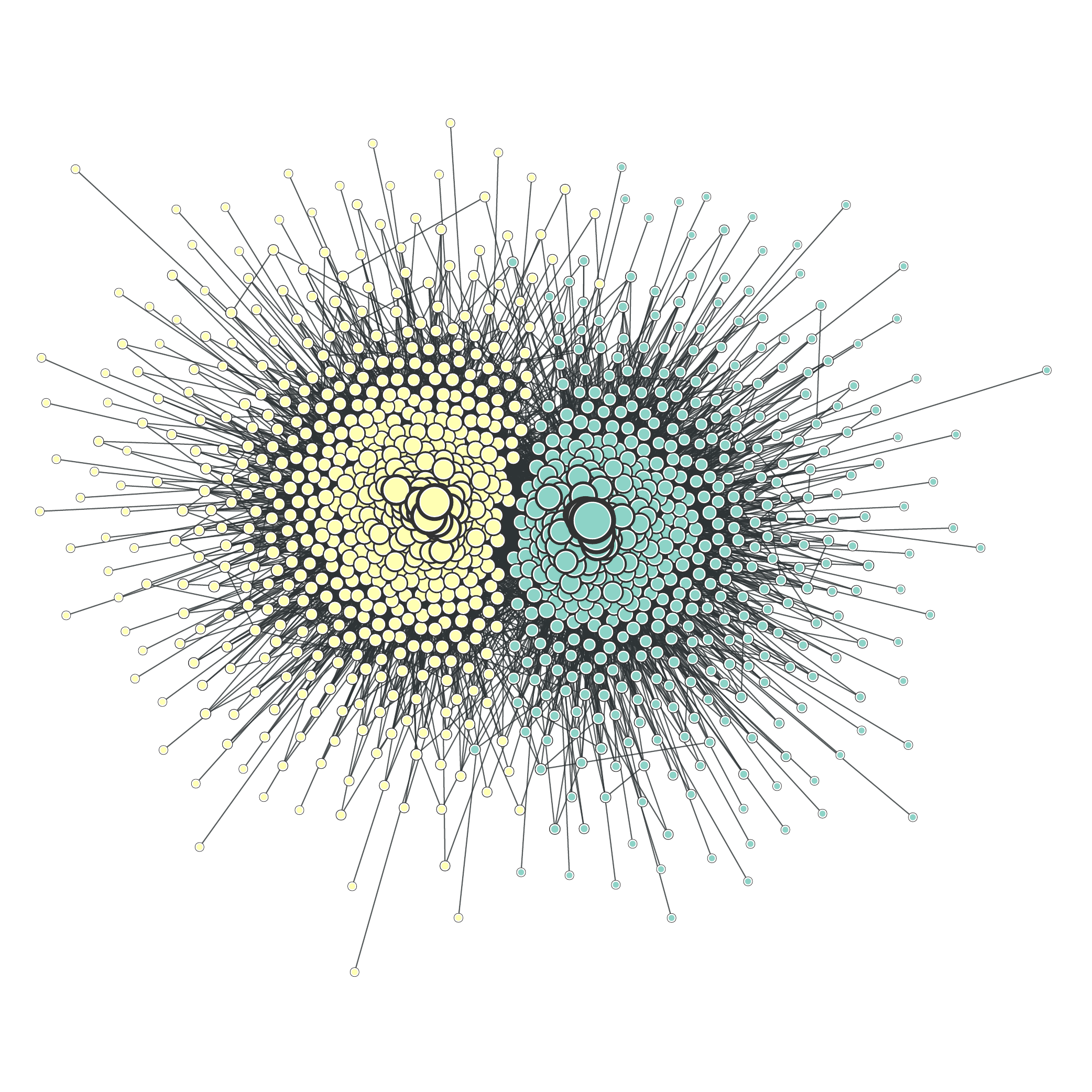}
      \put(0,0){(b)}
    \end{overpic}
  \end{tabular} \caption{Inferred groups for a political blog network
  generated form the maximum-entropy DC-SBM, inferred using the (a)
  Poisson DC-SBM and (b) the erased Poisson DC-SBM, with inferred groups
  indicated by the node colors, and node degrees by the node sizes. The
  layout was obtained with the spring-block algorithm of
  Ref.~\cite{hu_efficient_2005}, which tries to place nodes together if
  they are connected by an edge. The ``liberal'' and ``conservative''
  groups correspond to the visible left (yellow) and right (blue)
  clusters, respectively. The layout places nodes with high degree in
  the center of the figure, which in (a) are clustered into their own
  separate groups, whereas in (b) they are merged with their true
  category.
  \label{fig:polblogs}}
\end{figure*}

\begin{figure}
  \includegraphics[width=\columnwidth]{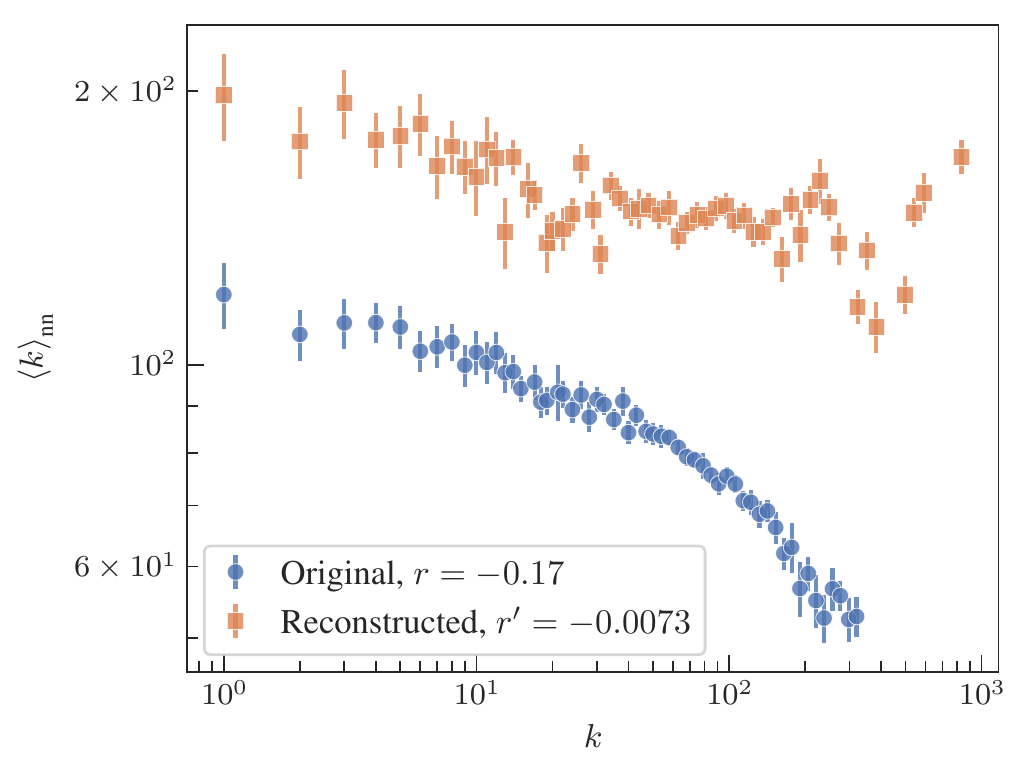}
  \caption{Mean degree of a neighbor of a node of degree $k$ as a
    function of $k$, for a political blog network generated form the
    maximum-entropy DC-SBM, and the corresponding inferred
    latent multigraph.\label{fig:polblogs-deg-corr}}
\end{figure}

In the following, we will compare two approaches to community detection:
1. Using the posterior $P(\bb|\A)$ based the Poisson multigraph model,
considering the simple graph observed as a possible instance; 2. Using the
posterior $P(\bb|\G)$ based on the erased Poisson model that generates
simple graphs exclusively. The original argument given by Karrer and
Newman~\cite{karrer_stochastic_2011} to justify the use of the former
approach is that for sparse graphs with an expected number of edges $E$
that is proportional to the number of nodes $N$, the mean parameter of
the Poisson distribution will decay as
\begin{equation}
  \frac{1}{N^2}\sum_{ij}\theta_i\theta_j\lambda_{b_ib_j} = \frac{2E}{N^2} = O(1/N).
\end{equation}
If we now consider the probability of observing more than one edge
between two nodes $i$ and $j$
\begin{multline}
  \sum_{A_{ij}=2}^{\infty}\frac{(\theta_i\theta_j\lambda_{b_ib_j})^{A_{ij}}\e^{-\theta_i\theta_j\lambda_{b_ib_j}}}{A_{ij}!}=\\
  1-\e^{-\theta_i\theta_j\lambda_{b_ib_j}}-\theta_i\theta_j\lambda_{b_ib_j}\e^{-\theta_i\theta_j\lambda_{b_ib_j}}=\\
  \frac{(\theta_i\theta_j\lambda_{b_ib_j})^2}{2} + O\left[(\theta_i\theta_j\lambda_{b_ib_j})^3\right],
\end{multline}
we can conclude that it will decay as $O(1/N^2)$ as long as the
corresponding parameters lie close to the mean, which itself decays as
$O(1/N)$. In this case the probability of observing multiple edges will
vanish for large $N$, and the model will generate mostly simple
graphs. The problem with this argument is that it breaks down precisely
when the network is heterogeneous and the parameters $\bm\theta$ and
$\bm\lambda$ are distributed with a high enough variance. In this case,
despite the vanishing value of the mean, we can in principle have a
sizable fraction of products $\theta_i\theta_j\lambda_{b_ib_j}$ that are
arbitrarily high. For example we could have this product approaching 1
for $N$ node pairs, and as along as the remaining $O(N^2)$ pairs decay
as $O(1/N)$, we still have the mean also scaling as $O(1/N)$, while the
resulting graph would have an abundance of multiedges, despite being
globally sparse. Ironically, this is precisely the situation one should
expect if the data possess a very strong community structure and very
broad degree distributions, making the Poisson model unsuitable. The
erased Poisson model, on the other hand, does not rely on uniform
sparsity, and should be able to better handle theses important
scenarios, which we investigate in the following.

\begin{figure*}
  \begin{tabular}{ccc}
    \begin{overpic}[width=0.33\textwidth]{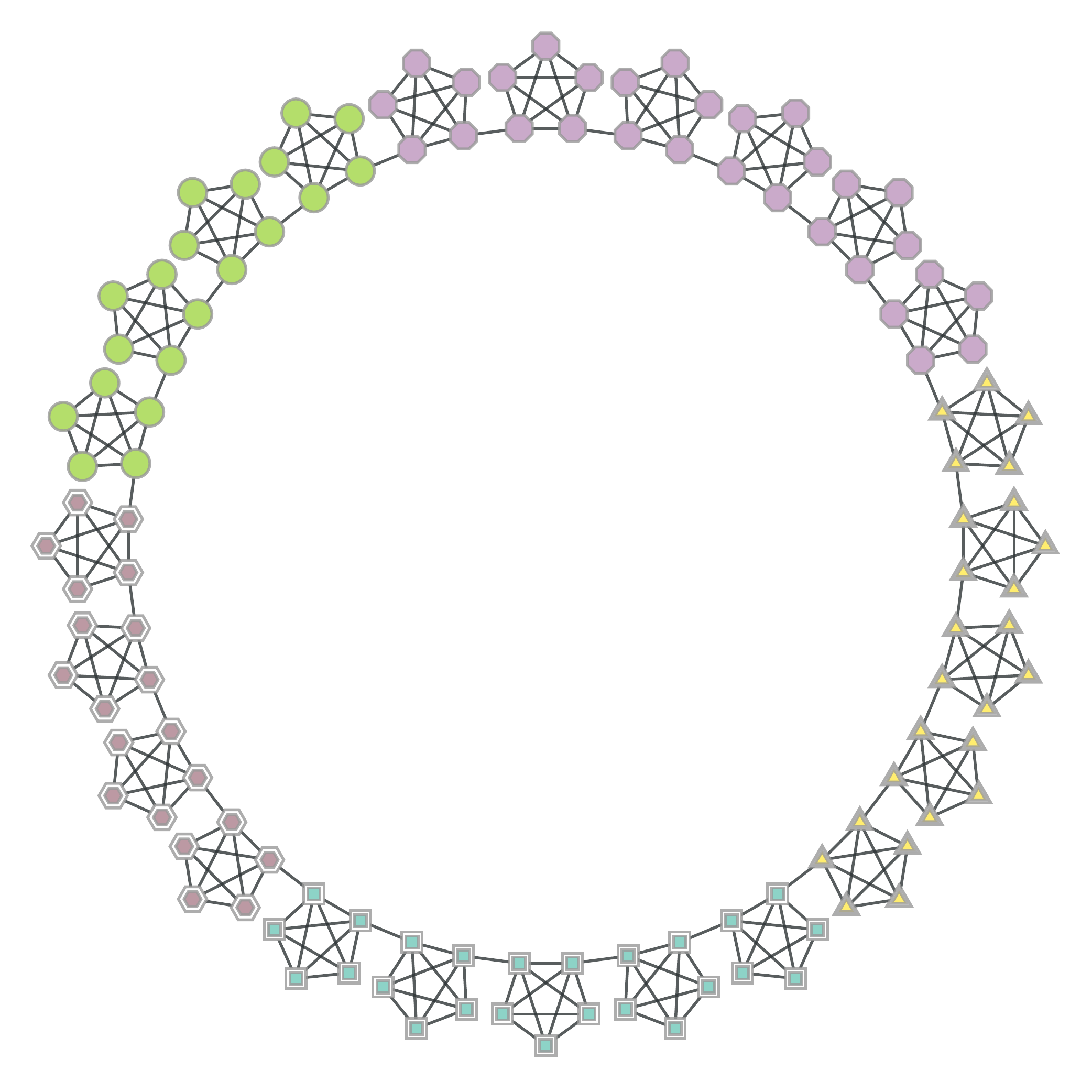}
      \put(0,0){(a)}
    \end{overpic}&
    \begin{overpic}[width=0.33\textwidth]{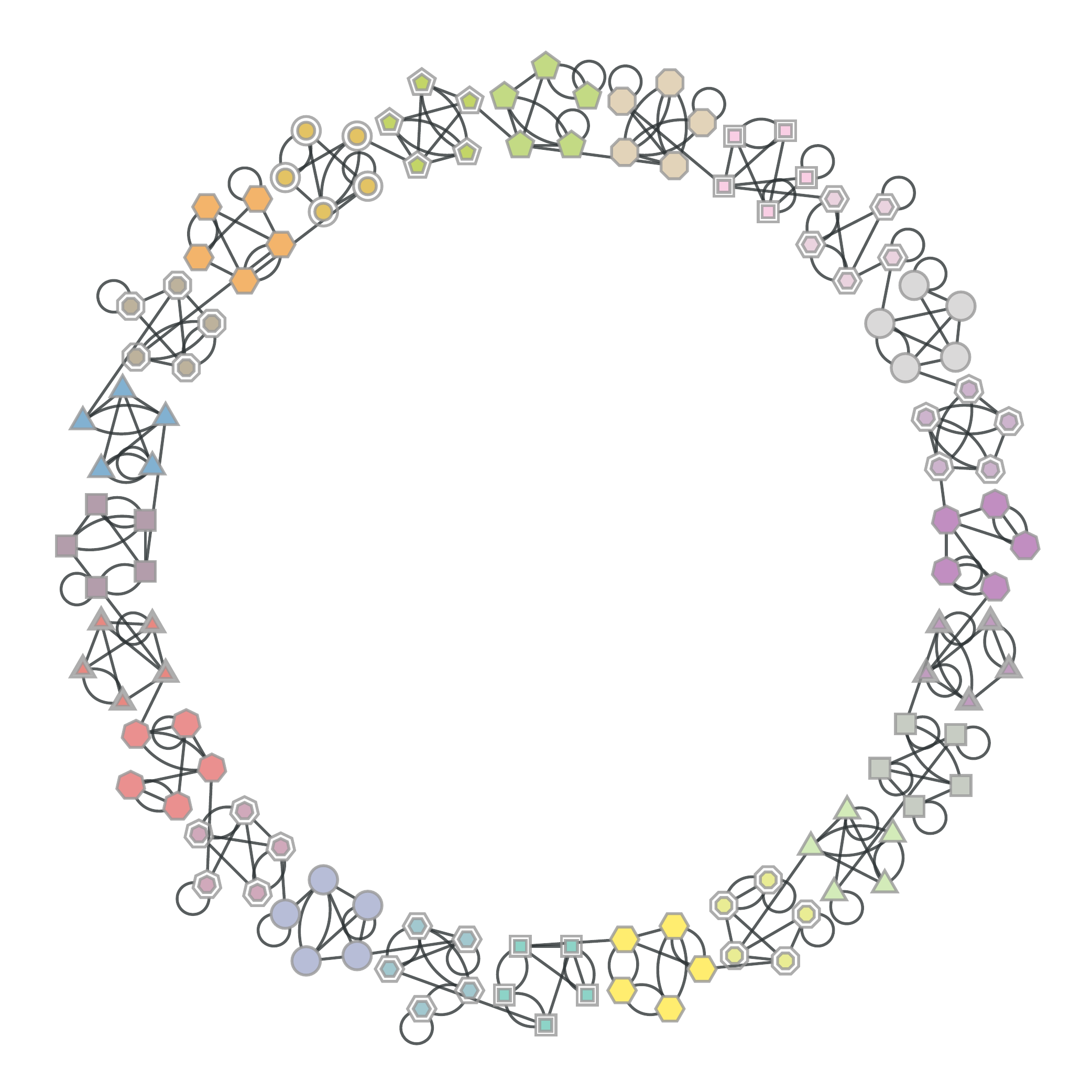}
      \put(0,0){(b)}
    \end{overpic}&
    \begin{overpic}[width=0.33\textwidth]{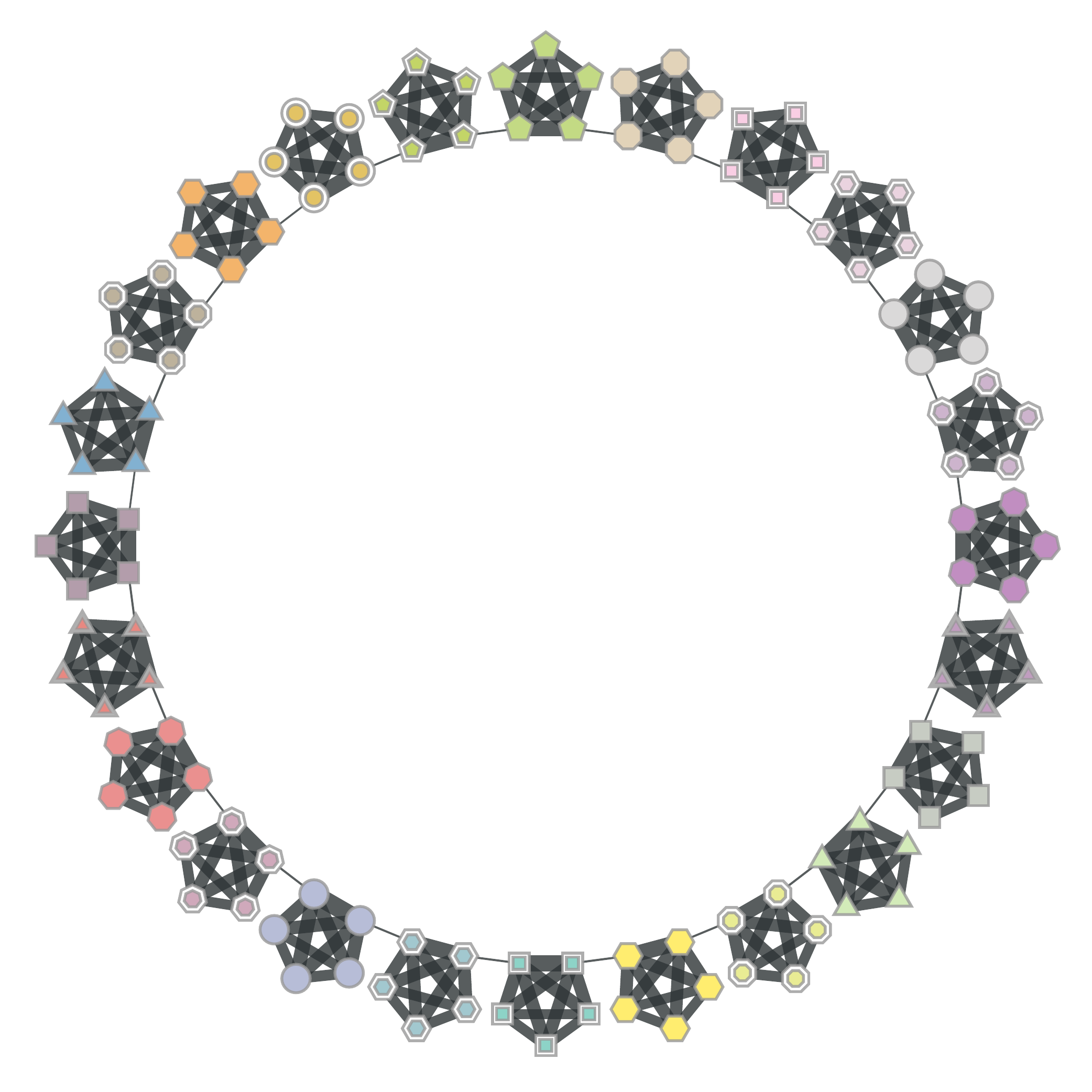}
      \put(0,0){(c)}
    \end{overpic}
  \end{tabular}

  \caption{(a) Network composed of a ring of 24 cliques of size 5,
  connected to each other by a single edge. The node colors and shapes
  correspond to a typical partition sampled from the posterior
  distribution of Eq.~\ref{eq:posterior}. (b) Network sampled from the
  maximum-likelihood Poisson DC-SBM obtained from the network in (a) and
  putting each clique in their own group (as shown by the node colors
  and shapes). (c) Fit of the erased Poisson nested DC-SBM to the
  network in (a), showing a sampled partition from the posterior
  distribution (node shape and color), coinciding perfectly with the
  individual cliques, and the marginal posterior distribution of edge
  multiplicities (edge thickness).\label{fig:cliques}}
\end{figure*}

\begin{figure}
  \begin{tabular}{cc}
    \begin{overpic}[width=.5\columnwidth]{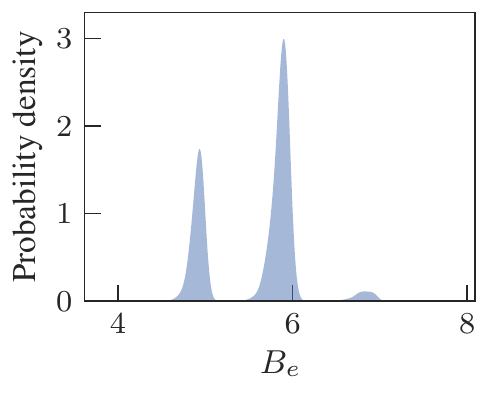}
      \put(0,0){(a)}
    \end{overpic} &
    \begin{overpic}[width=.5\columnwidth]{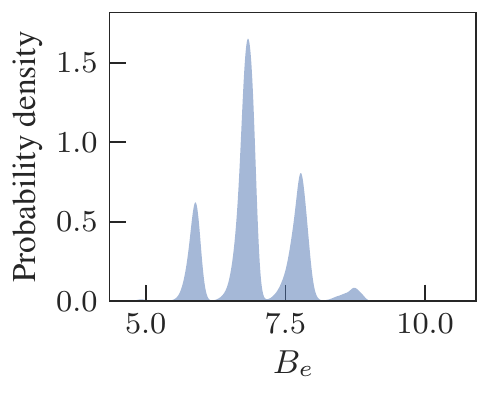}
      \put(0,0){(b)}
    \end{overpic}\\
    \begin{overpic}[width=.5\columnwidth]{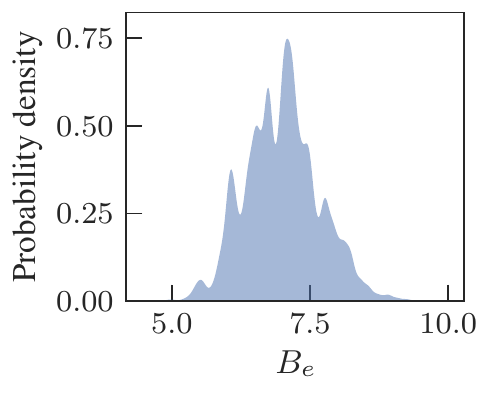}
      \put(0,0){(c)}
    \end{overpic}&
    \begin{overpic}[width=.5\columnwidth]{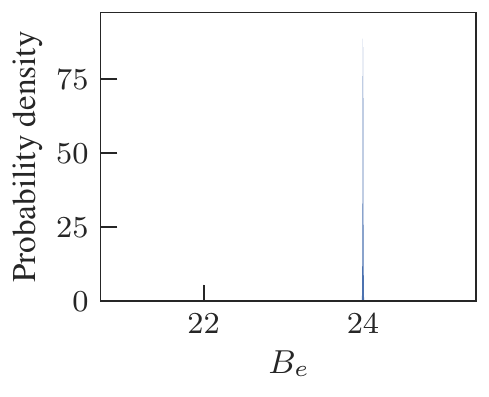}
      \put(0,0){(d)}
    \end{overpic}
  \end{tabular}

  \caption{Posterior distribution of effective number of groups $B_e$
  for the network in Fig.~\ref{fig:cliques}a, obtained with (a) the Poisson
  DC-SBM, (b) the nested Poisson DC-SBM, (c) the erased Poisson DC-SBM,
  and (d) the nested erased Poisson DC-SBM, the latter showing a
  distribution highly concentrated on
  $B_e=24$.\label{fig:cliques_posterior}}
\end{figure}

\begin{figure}
  \includegraphics[width=\columnwidth]{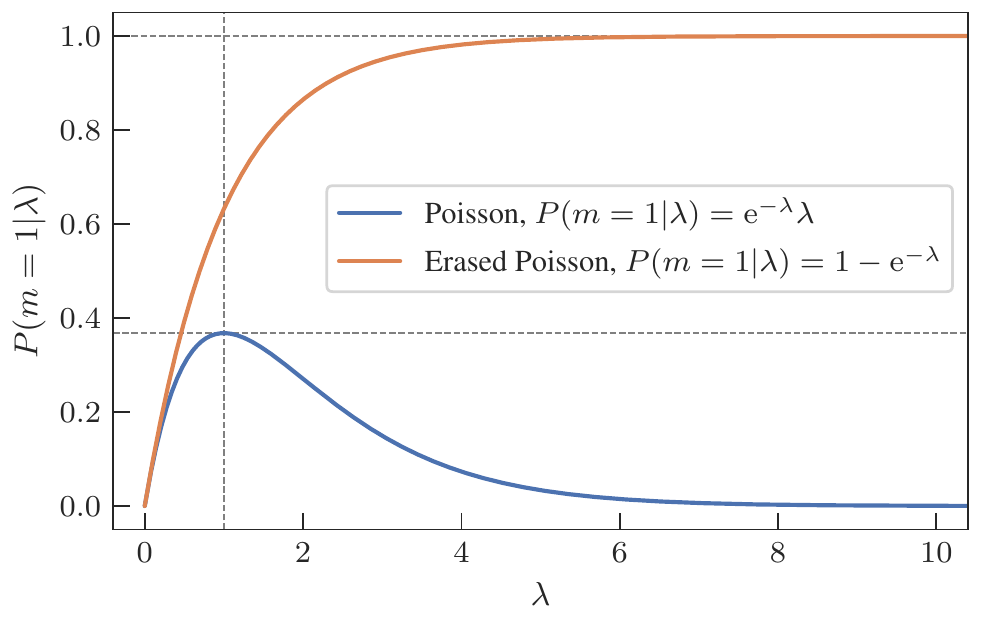}
  \caption{Probability of observing a sample $m=1$ from the Poisson and
  erased Poisson models, as a function of the parameter $\lambda$, as
  shown in the legend. The vertical and horizontal lines show the
  maximum of the Poisson at $\lambda=1$ and $P(m=1|\lambda)=1/\e$, and
  the asymptotic value of $P(m=1|\lambda)\to 1$ for the erased Poisson
  model as $\lambda\to\infty$. \label{fig:poisson-simple}}
\end{figure}

\begin{figure}
  \begin{tabular}{cc}
    \multicolumn{2}{c}{Karate club}\\
    \begin{overpic}[width=.5\columnwidth, trim=0 -1cm 0 0]{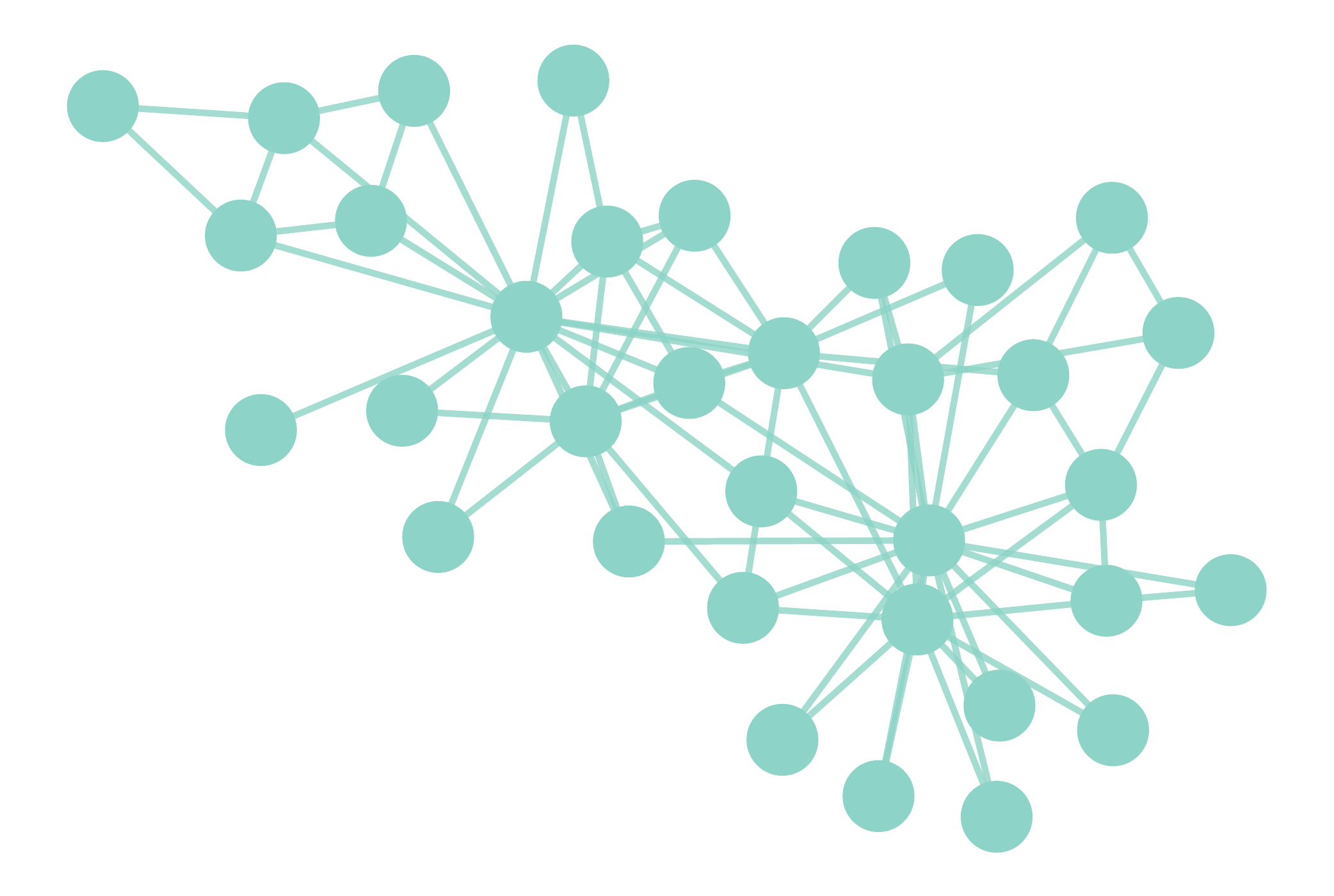}
      \put(40,0){\smaller Poisson}
    \end{overpic}&
    \begin{overpic}[width=.5\columnwidth, trim=0 -1cm 0 0]{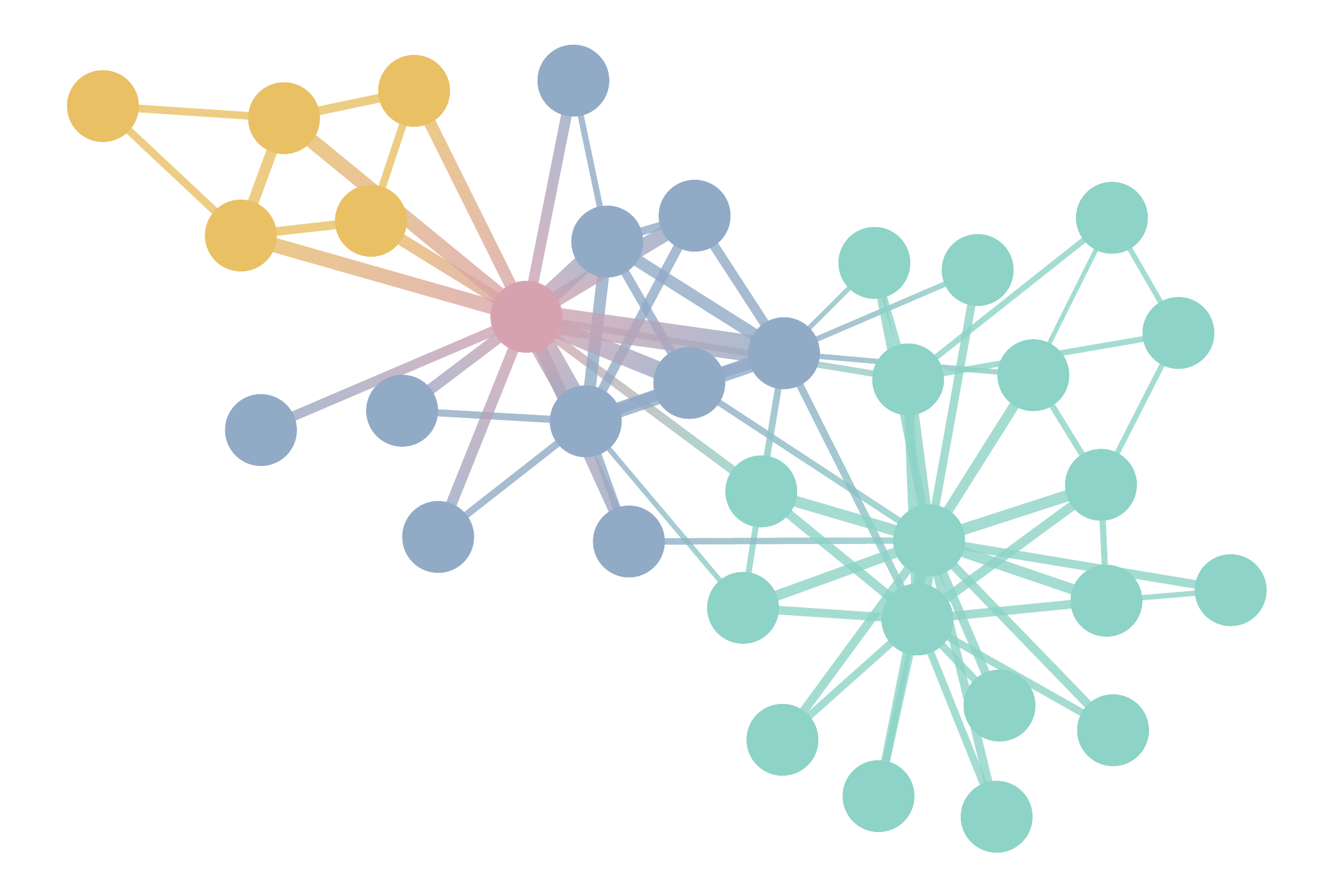}
      \put(25,0){\smaller Erased Poisson}
    \end{overpic}\\
    \multicolumn{2}{c}{Dolphins}\\
    \begin{overpic}[width=.5\columnwidth, trim=0 -1cm 0 0]{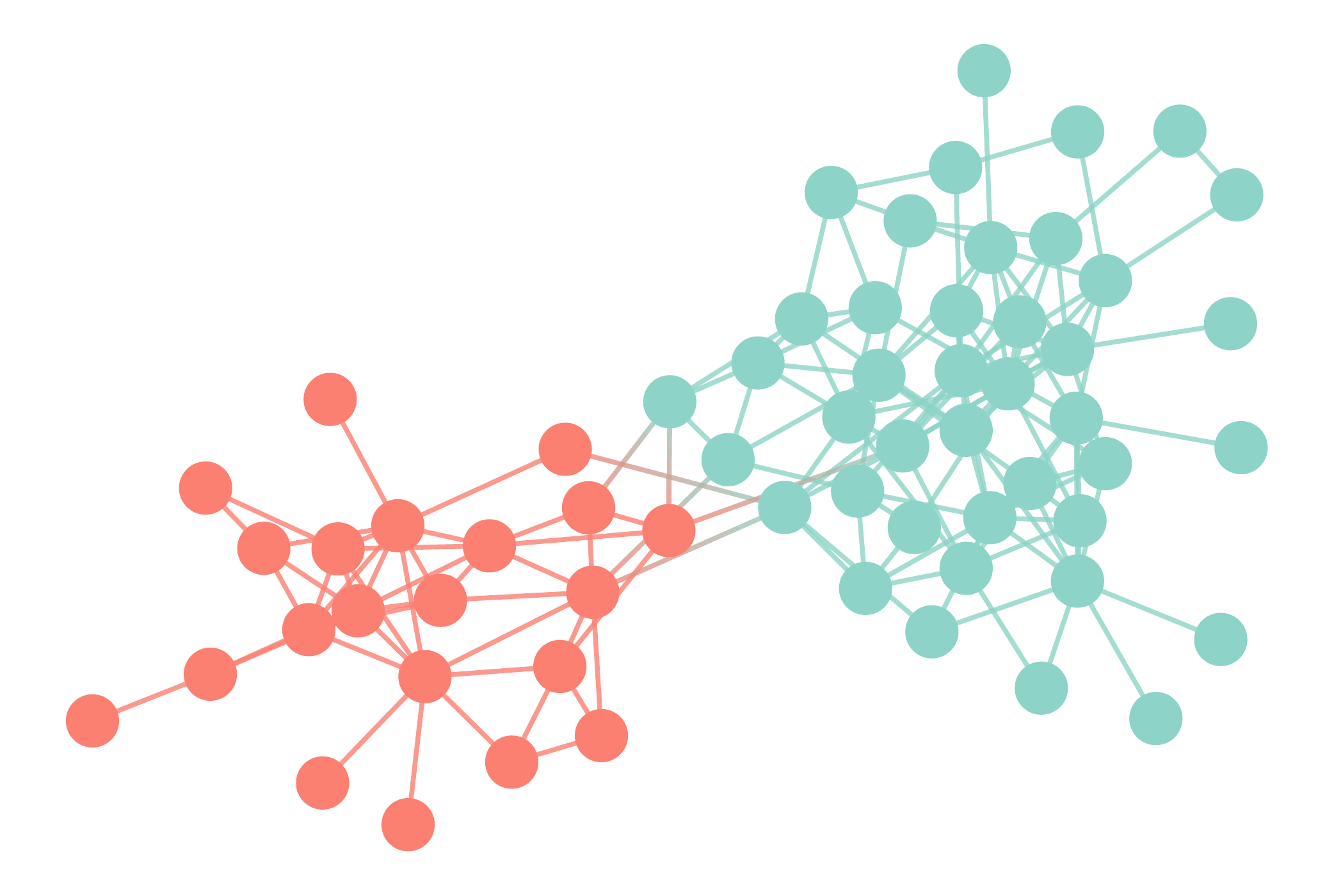}
      \put(40,0){\smaller Poisson}
    \end{overpic}&
    \begin{overpic}[width=.5\columnwidth, trim=0 -1cm 0 0]{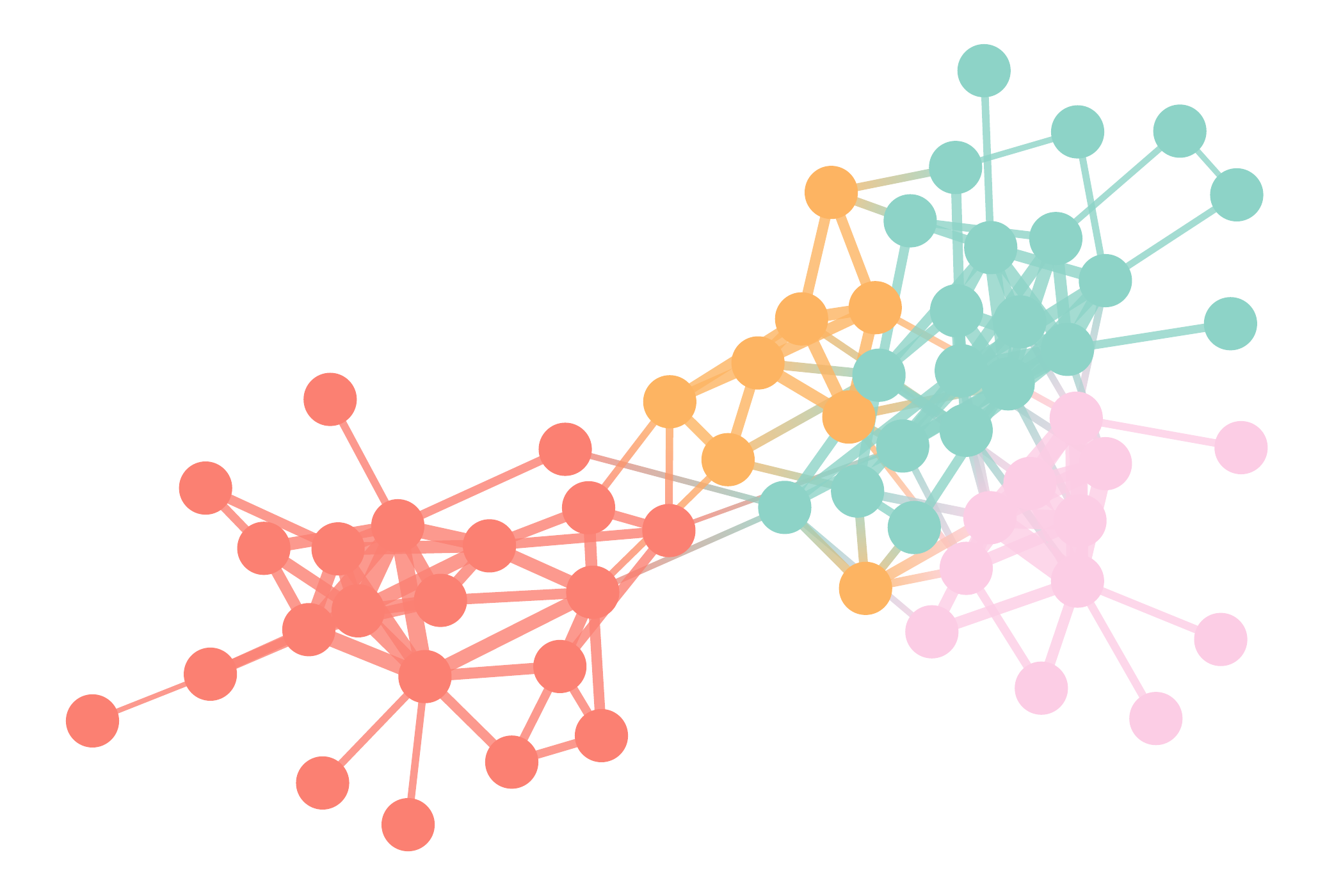}
      \put(25,0){Erased Poisson}
    \end{overpic}\\
    \multicolumn{2}{c}{Political books}\\
    \begin{overpic}[width=.5\columnwidth, trim=0 -1cm 0 0]{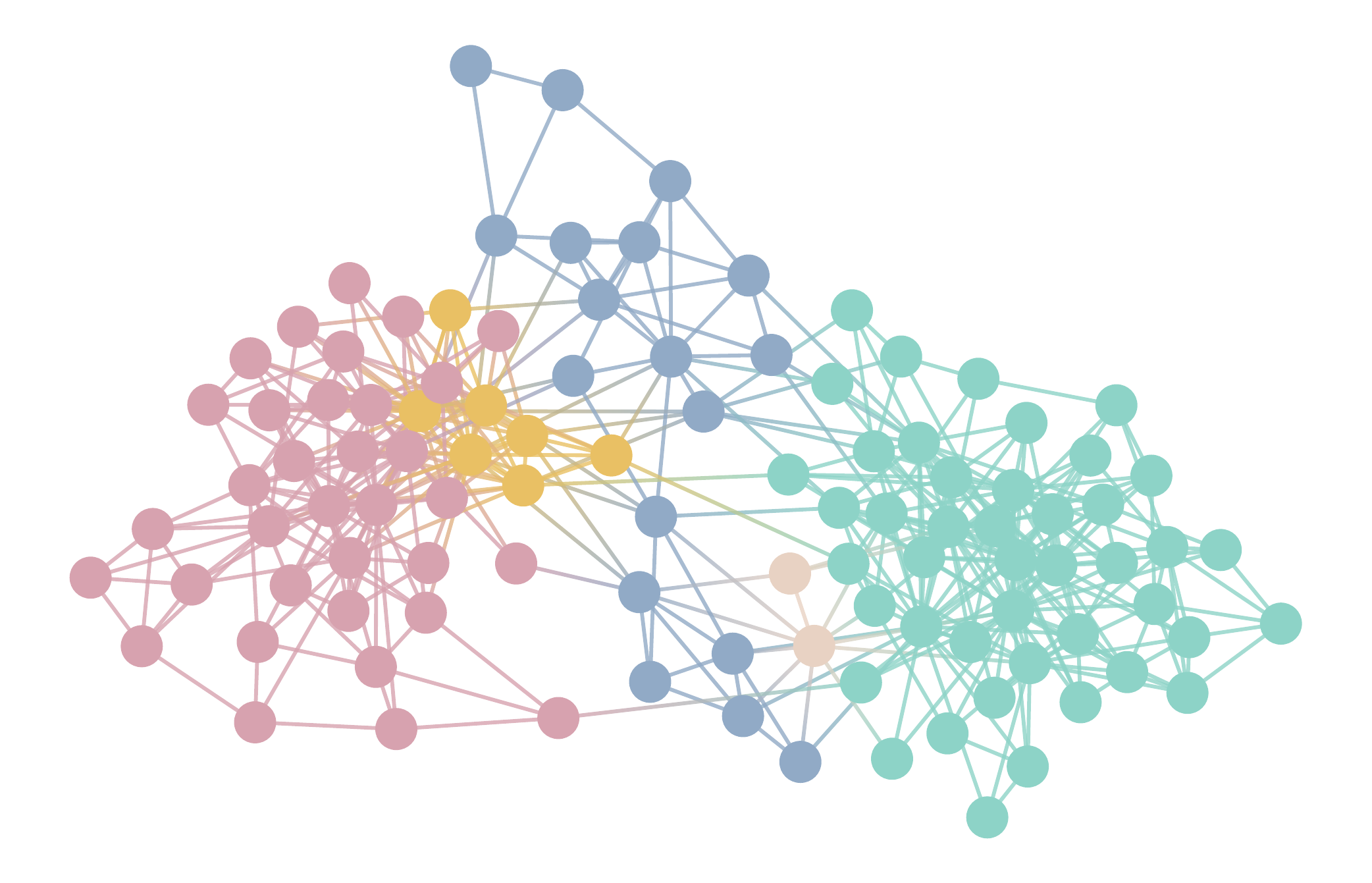}
      \put(40,0){\smaller Poisson}
    \end{overpic}&
    \begin{overpic}[width=.5\columnwidth, trim=0 -1cm 0 0]{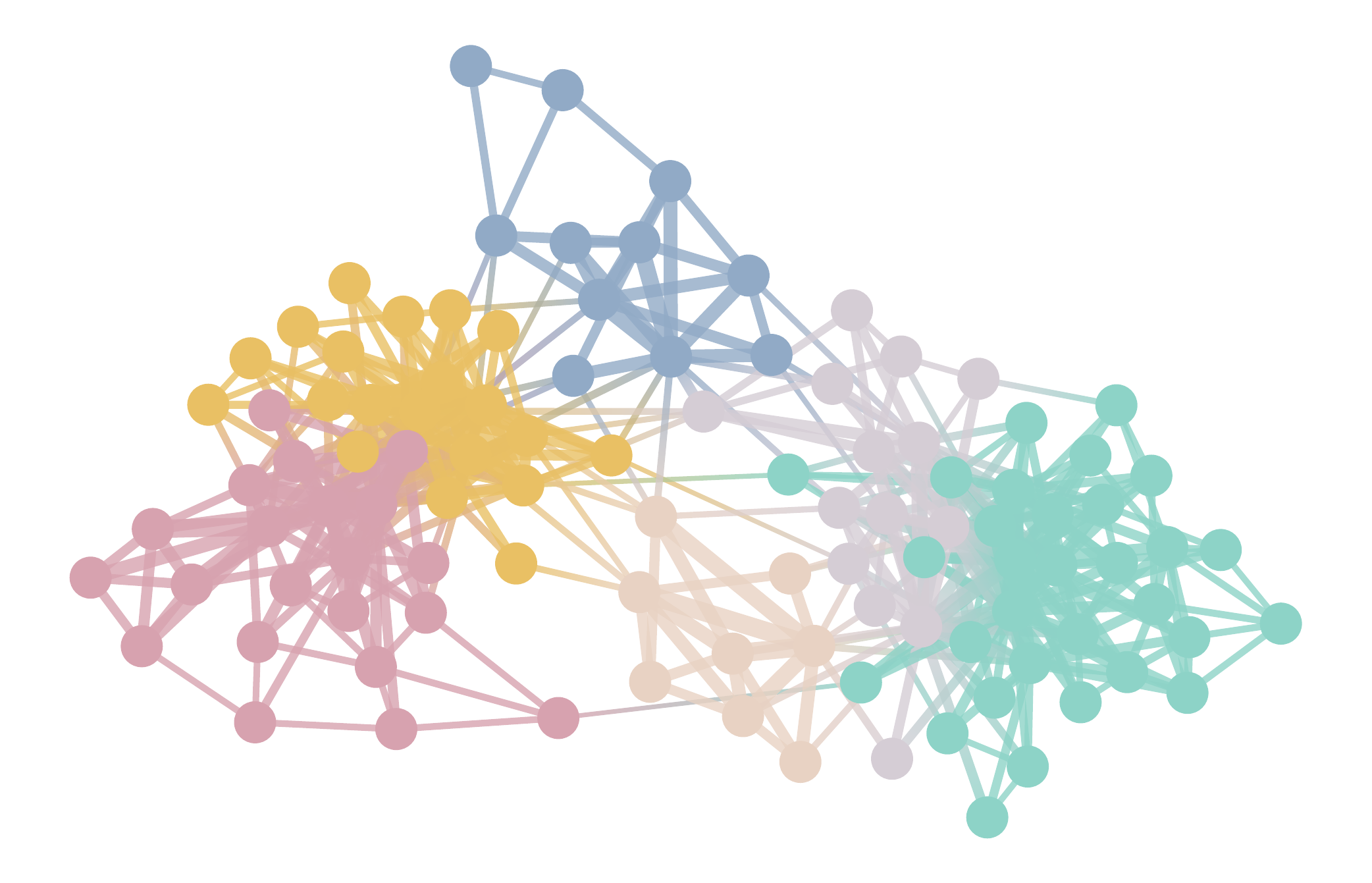}
      \put(25,0){\smaller Erased Poisson}
    \end{overpic}
  \end{tabular}

  \caption{Inferred group memberships for Zachary's karate club
  network~\cite{zachary_information_1977}, a dolphin social
  network~\cite{lusseau_bottlenose_2003}, and co-purchases of political
  books~\cite{krebs_political_nodate}, using the Poisson DC-SBM and the
  erased Poisson DC-SBM. In each network, the latter model reveals a
  larger number of groups, due to its increased ability of identifying
  heterogeneous densities. \label{fig:small_nets}}
\end{figure}

\subsection{Broad degree distributions}

We begin illustrating the behavior or the erased Poisson model with the
network of political blogs of Adamic and
Glance~\cite{adamic_political_2005}, which describes the citations
between blogs during the 2004 US elections. Either version of the model
finds a wealth of information, dividing the network in many groups. In
order to simplify our analysis, we use the known division between
liberal and conservative blogs as an imposed partition of randomly
generated networks, which we sample from the maximum-entropy DC-SBM of
Eq.~\ref{eq:sbm_maxent} that preserves the number of edges that go
between nodes of the same and different groups, as well as the node
degrees, when compared to the real network. This means that, in our
analysis, this division is indeed the true one, instead of only
putatively true, as is the case of the empirical network. If we now
employ the Poisson DC-SBM to the resulting network, we get the partition
into five groups, as shown in Fig.~\ref{fig:polblogs}a. Since the
degrees constraints induce disassortative degree-degree correlations,
something that is not expected with the Poisson model, the inference of
that model attempts to account for this pattern by subdividing each
group into subgroups of nodes with similar degrees, in an attempt to
account for this existing mixing pattern as specific probabilities of
connections between these extra groups. If instead we use the erased
Poisson model, we uncover correctly only the two planted partitions, as
we see in Fig.~\ref{fig:polblogs}b, as this model is capable of
incorporating the induced degree disassortativity intrinsically. Indeed,
by inspecting the inferred latent multigraph, we see that it lacks a
substantial fraction of the disassortativity, as shown in
Fig.~\ref{fig:polblogs-deg-corr}b, which then only emerges once the
multiedges are erased.

\subsection{Heterogeneous densities}

We turn now to a related, but different scenario where the Poisson model
also gives suboptimal results. We consider the artificial network
composed of a ring of 24 cliques of size 5. This kind of network was
used as an example by Good et al~\cite{good_performance_2010} of a
situation where community detection methods fail to find the more
obvious pattern. Indeed, as was shown recently by Riolo and
Newman~\cite{riolo_consistency_2019}, inferring the DC-SBM also yields
unsatisfactory results, where adjacent cliques are merged together, as
shown in Fig.~\ref{fig:cliques}a. In Fig.~\ref{fig:cliques_posterior}a
we see the posterior distribution of effective number of groups, defined
as $B_{\text{e}}=\e^{S_{\text{e}}}$, with
\begin{equation}
  S_{\text{e}} = -\sum_r \frac{n_r}{N}\ln\frac{n_r}{N},
\end{equation}
being the entropy of the membership distribution. For the DC-SBM above,
the posterior distribution fluctuates around 6 groups, falling
significantly short of the expected 24.

At first, one might think that this problem is due to an underfitting of
the model, caused by the use of noninformative priors. As was shown in
Refs.~\cite{peixoto_parsimonious_2013,peixoto_nonparametric_2017}, the
use of such priors incurs a penalty in the posterior log-probability
that grows quadratically with the number of groups, which in turn means
that no more than $O(\sqrt{N})$ groups can ever be inferred in sparse
networks. This issue is resolved by replacing the noninformative priors
by a deeper hierarchy of priors and their hyperpriors, forming a nested
DC-SBM~\cite{peixoto_hierarchical_2014}. The use of the nested model,
which remains nonparametric and agnostic about mixing patterns,
increases the inference resolution by enabling the identification of up
to $O(N/\log N)$ groups. In a similar example of a network composed of
64 cliques of size 10, the nested model is capable of identifying all 64
cliques, whereas the ``flat'' version finds only 32 groups, composed each
of 2 cliques~\cite{peixoto_bayesian_2019}.

However, when applied to the current example, the use of the nested
Poisson DC-SBM is not sufficient to uncover all 24 cliques. In
Fig.~\ref{fig:cliques_posterior}b we see the posterior distribution of
effective group sizes for the nested Poisson model. Although the mean
number of groups increases, it still falls short of the 24. This
indicates that the problem may be not only underfitting, but also
mispecification,
i.e. the model is simply not adequate to describe the data. Indeed a
closer inspection reveals that this is precisely the case. A version of
the DC-SBM that should be able to generate the given number of cliques
would be one where the probability of an edge existing between two nodes
of the same clique would be very close to one. However, the Poisson
model struggles at describing this structure because it cannot allow for
a single edge occurring with such a high probability, without generating
multiple edges as well. As is illustrated in
Fig.~\ref{fig:poisson-simple}, the Poisson distribution can generate the
occurrence of a single edge with a probability at most $1/\e \approx
0.368$, and even in that case the occurrence of multiple edges is no
longer negligible. This limitation is absent from the erased Poisson
model, which can describe arbitrary probabilities of single edges. In
Fig.~\ref{fig:cliques}b we show a sample of the DC-SBM with parameters
chosen so that it replicates the original network as closely as possible:
the nodes are separated into 24 groups of size 5, the mean number of
edges between nodes of the same group is one, and between adjacent
groups is 1/25. The resulting network is not only riddled with
multiedges and self-loops, but its also shows a far more irregular
structure than the original one. However, through the lenses of the
Poisson model, both networks are difficult to distinguish, as they have
very similar likelihoods. Although it can be possible to extract useful
information even from mispecified models via a detailed inspection of
the posterior distribution~\cite{riolo_consistency_2019}
--- a powerful feature of Bayesian methods --- this is not a satisfying
resolution for such a simple example. Indeed, what we have is once more
a situation where the network is globally sparse but locally dense, and
we should expect the erased Poisson model to behave better. In fact,
using the nested DC-SBM based on the erased Poisson model we can uncover
all 24 cliques, as it is capable of describing their probability more
accurately. In Fig.~\ref{fig:cliques}c is shown the result obtained with
the erased Poisson model, together with the inferred latent multigraph,
which has a abundance of multiedges inside each group, that translate
into cliques once erased, with the probability of each edge approaching
one. Importantly, the successful detection of the cliques is only
possible if the erased Poisson model is used together with the nested
priors of Ref.~\cite{peixoto_nonparametric_2017}, which illustrates the
combined effect of more appropriate model specification with structured
priors that prevent underfitting.

We further illustrate the use of the erased Poisson model with some
further empirical examples in Fig.~\ref{fig:small_nets}, comprised of a
social network between members of a karate
club~\cite{zachary_information_1977}, an animal social network between
bottlenose dolphins~\cite{lusseau_bottlenose_2003}, and co-purchases of
books about american politics~\cite{krebs_political_nodate}. In each
case, using the erased Poisson DC-SBM we obtain a more detailed division
of the network, with a larger number of groups, when compared to the
employing the Poisson DC-SBM directly. The most extreme difference is
obtained by the karate club network, where the Poisson DC-SBM yields a
single group, but the erased Poisson version yields four groups. The
explanation for the difference in each case is the same as for the
ring-of-cliques example considered previously: since the Poisson DC-SBM
is unable to ascribe high probabilities to the existence of edges, it
puts a smaller statistical weight to dense regions of the network, even
when that would be sufficient to point to the existence of a separate
group. The erased Poisson model does not have this limitation and hence
is able to isolate this kind of structure with more confidence.

\subsection{Modularity and group assortativity}

An important pattern in network structure is the degree of
assortativity, or homophily, between node types. This is commonly
measured via the modularity quantity~\cite{newman_finding_2004}, which
counts the excess of edges between nodes of the same type, when compared
to a null model without any homophily, i.e.
\begin{equation}
  Q = \frac{1}{2E}\sum_{i\neq j}\left(G_{ij} - \avg{G_{ij}}\right)\delta_{b_i,b_j},
\end{equation}
where $\avg{G_{ij}}$ is the expectation of an edge $(i,j)$ existing
according to the chosen null model, and the normalization guarantees
$Q\in[-1,1]$. The most often used null model is
$\avg{G_{ij}}=k_ik_j/2E$, which corresponds to a Poisson multigraph
model with a maximum-likelihood choice of fugacities
$\theta_i=k_i/\sqrt{2E}$. As we have discussed, the Poisson model
approaches the maximum-entropy simple graph model if the degrees are
sufficiently smaller than $\sqrt{2E}$, otherwise this assumption becomes
inadequate to describe null models of simple graphs. We can use the
erased Poisson model as a better alternative in two different ways, the
first of which is by simply using its expected value
$\avg{G_{ij}}=1-\e^{-\theta_i\theta_j}$, which yields
\begin{align}
  Q &= \frac{1}{2E}\sum_{i\neq j}\left[G_{ij} - (1-\e^{-\theta_i\theta_j})\right]\delta_{b_i,b_j},\\
    &= \frac{1}{2E}\left[\sum_re_{rr} - n_r(n_r-1) + \sum_{i\neq j}\e^{-\theta_i\theta_j}\delta_{b_i,b_j}\right],
\end{align}
with $2E=\sum_{ij}G_{ij}$.  The values of $\bm\theta$ can be obtained
efficiently with the EM algorithm presented in
Sec~\ref{sec:reconstruction}. A disadvantage of this approach is that
the computation of the last term in the above equation requires
$O[(N/B)^2]$ operations, and thus is not very efficient for large
networks. The second approach we describe is faster, and is comprised of
the computation of modularity for the multigraph inferred from
$P(\A|\G,\hat{\bm\theta})$ using the same EM algorithm (instead of the
simple graph $\G$ directly), for which $\avg{A_{ij}}=\theta_i\theta_j$
becomes the appropriate null model, i.e.
\begin{align}
  Q &= \frac{1}{2E'}\sum_{ij}\left(w_{ij} - \theta_i\theta_j\right)\delta_{b_i,b_j},\\
  &= \frac{1}{2E'}\sum_r\omega_{rr} - \hat\theta_r^2 \\
  &= \frac{1}{2E'}\sum_r\omega_{rr} - \frac{\omega_r^2}{2E'}
\end{align}
with with $2E'=\sum_{ij}w_{ij}$,
$\omega_{rs}=\sum_{ij}w_{ij}\delta_{b_i,r}\delta_{b_j,s}$,
$\omega_r=\sum_s\omega_{rs}$, $\hat\theta_r=\sum_i\theta_i\delta_{b_i,r}$,
and $\theta_i=\sum_jw_{ji}/\sqrt{2E'}$. This quantity can be computed in
time $O(E+N)$, and thus offers a significant speed advantage over the
first one. The two approaches are not identical, and we should not
expect to obtain the same value of $Q$ between them in general, but in
case the network was in fact sampled from the erased Poisson model, we
must have $Q\approx 0$ with either computation.

We note that the use of modularity maximization with the purpose of
identifying communities in networks, although a popular approach, is
ill-advised. This is because that method cannot account for the
statistical significance of the node partitions found, and can lead to
misleading results, such as high-scoring partitions in fully random
graphs~\cite{guimera_modularity_2004}, non-modular networks such as
trees~\cite{bagrow_communities_2012}, has been shown to systematically
overfit empirical data~\cite{ghasemian_evaluating_2019}, while at the
same time it will fail for networks with obvious community
structure~\cite{fortunato_resolution_2007,good_performance_2010}. Nevertheless,
if the partitions are obtained with some other method (like the one of
we have described in the previous section, which suffers from none of
the mentioned shortcomings), or originate from network annotations, the
value of $Q$ can be a good description of the existing homophily, and
the corrections above can be used to improve it.

\section{Conclusion}\label{sec:conclusion}

We have considered the use of the erased Poisson model to describe
simple graphs with different kinds of heterogeneous sparsity, in
particular with broad degree distributions and community structure. We
have shown how this model can give rise to intrinsic degree-degree
correlations that are very similar to those existing in maximum-entropy
models of simple graphs. We have presented an expectation-maximization
(EM) algorithm to infer the underlying Poisson model from simple graph
data, and shown how it can be used to potentially explain observed
disassortative degree-degree correlations, if they arise predominantly
from the imposed degrees. Previously, this could only have been
determined by generating networks from an appropriate null model, and
comparing the assortativity obtained. Our approach is more constructive,
since it yields an inference of a generative model, rather than simply a
comparison with a null one. This means it is more informative in
situations where the degree constraints can account for only a portion
of the correlations observed, in which case our approach yields a
residual multigraph, with a subtracted contribution of the degree
constraints to the degree correlations, which can be further analyzed in
arbitrary ways.

We have also investigated the use of this model in community detection,
and shown how it is more adequate to uncover communities not only in
simple graphs with broad degree distributions, but also when they
possess strong community structure. In the latter case, the erased
Poisson model is capable of combining degree correction with the
existence of edge probabilities approaching one, meaning it can easily
model networks that are globally sparse, but locally dense. The enhanced
explanatory power is achieved by sacrificing neither mathematical
tractability nor algorithmic efficiency.

The erased Poisson model has been used before as a means to combine
multigraph generative models with measurement models for simple graphs,
when performing joint network reconstruction with community detection in
Refs.~\cite{peixoto_reconstructing_2018,peixoto_network_2019}, although
these works omitted a detailed analysis of this modelling
approach. Since the erased Poisson model is better specified for
networks with strongly heterogeneous density, it remains to be
determined to what extent it can improve link prediction and network
reconstruction, when compared to alternatives. We leave this
investigation for future work.

\appendix

\section{Ensemble equivalences}\label{app:equivalence}

In this section we consider network generative processes that at first
might seem distinct, but in fact are equivalent not only to each other
but also to the Poisson model considered in the main text.

\subsection{Sequential edge-dropping model}

We consider the situation where a random multigraph is grown by adding
$E$ edges one by one to the network in sequence, and the probability
that a given edge is placed between nodes $i$ and $j$ is given by
$q_{ij}$, with $\sum_{i<j}q_{ij} = 1$. In this case, the probability of
observing a final multigraph $\A$ is given by a multinomial distribution
\begin{equation}
  P(\A|E) = E!\prod_{i<j}\frac{q_{ij}^{A_{ij}}}{A_{ij}!}.
\end{equation}
Now if the total number of edges is also allowed to vary, and it is
first sampled from a Poisson distribution with mean $\lambda$,
$P(E)=\lambda^E\e^{-\lambda}/E!$, we have that the marginal probability
will be a product of independent Poisson distributions
\begin{align}
  P(\A) &= \sum_EP(\A|E)P(E) \\
        &= \prod_{i<j}\frac{(\lambda q_{ij})^{A_{ij}}\e^{-\lambda q_{ij}}}{A_{ij}!}.\label{eq:edge_dropping}
\end{align}
If we make the choice $q_{ij} = \theta_i\theta_j / \lambda$ and
$\lambda=\sum_{i<j}\theta_i\theta_j$ we recover the Poisson model of
Eq.~\ref{eq:poisson}, and likewise allowing for self-loops we recover
Eq.~\ref{eq:poisson-sl}.

This ``edge-dropping'' process is a simple model of a growing network
where the placement of new edges is not affected by the existing
edges. While this assumption is likely to be violated in a variety of
more realistic settings, the central point here is to notice that it
implicitly assumes a distinguishability of the multiedges, due to the
order in which they appear. Therefore, a maximum-entropy model that
assumes edge distinguishability is the appropriate null model when edges
are sampled individually.

\subsection{Microcanonical configuration model}\label{sec:microcanonical}

The configuration or ``stub matching'' model is a standard procedure for
generating multigraphs with prescribed degree
sequences~\cite{bollobas_probabilistic_1980,fosdick_configuring_2018}:
to each node $i$ is attributed a number $k_i$ of distinguishable
``stubs'' or ``half-edges'', which are then paired uniformly at random,
allowing for multiedges and self-loops. Since every pairing --- or
``configuration'' --- occurs with the same probability, this is a
maximum-entropy microcanonical ensemble of configurations (rather than
multigraphs), with the prescribed degree sequence functioning as a
constraint. This is different from the ``canonical'' ensembles we have
been considering so far, where the degrees are constrained only in
expectation. Although the configurations are uniformly distributed, the
associated multigraphs are not, since more than one configuration will
map to the same multigraph. We can obtain the probability of observing a
particular multigraph by enumerating the corresponding
configurations. With $2E=\sum_ik_i$ half-edges, we can count the total
number of configurations by starting with any arbitrary half-edge, which
can then be paired with $2E-1$ other half-edges. For any of these
choices, we can pick any of the remaining half-edges which can be paired
with any of the other remaining $2E-3$ ones. Proceeding in this way we
have that the total number of pairings is $(2E - 1) \times (2E-3) \times
(2E -
5) \times \cdots \times 1 = (2E-1)!!$. To account for multigraphs, we
observe that for each node with $k_i$ half-edges
there are $k_i!$ permutations of their matchings that yield different
configurations but correspond to the same multigraph, if all matched
half-edges belong to different nodes. Otherwise, this over-counts
$A_{ij}!$ label permutations of half-edges matched between nodes $i$ and
$j$, and likewise $A_{ii}!!$ permutations for self-loops matched for the
same node. Putting all this together, we have that the
multigraphs are distributed according to the ratio
\begin{equation}
  P(\A | \bm k) = \frac{\prod_ik_i!}{(2E-1)!!\prod_{i<j}A_{ij}!\prod_iA_{ii}!!},
\end{equation}
assuming $\sum_jA_{ij} = k_i$ for every node $i$, otherwise $P(\A |\bm
k) = 0$. We note that all generated graphs that happen to be simple with
$A_{ij}\in\{0,1\}$ occur with the same probability
$\prod_ik_i!/(2E-1)!!$. Therefore if we discard all multigraphs, the
resulting simple graph ensemble has maximum-entropy (but with a new
normalization constant that is intractable in
general~\cite{bender_asymptotic_1978}, and even in simpler cases where
all $k_i$ are equal~\cite{wormald_models_1999}).

For an arbitrary (but in this case necessarily integer) choice of
$k_i=\hat{k}_i$, this microcanonical ensemble is not equivalent to any
of the previous canonical ones, since those allow for fluctuations of
the degrees around its imposed expected value. Indeed, this lack of
ensemble equivalence persists even in the limit
$N\to\infty$~\cite{anand_gibbs_2010,squartini_breaking_2015}, unlike
more typical situations where the number of imposed constraints is
fixed. In the latter case, asymptotic equivalence between ensembles is
expected, but since the number of constraints given by
Eq.~\ref{eq:const_k} is extensive, i.e. grows with $N$, this equivalence
is never realized.\footnote{Strictly speaking, the results of
Refs.~\cite{anand_gibbs_2010,squartini_breaking_2015} refer to maximum
entropy ensembles of simple graphs with hard and soft constraints, but
the main arguments are also valid for the configuration and Poisson
models.}

In spite of the lack of asymptotic equivalence, an exact equivalence
with the Poisson model does exist once we consider the degrees $k_i$ and
fugacities $\theta_i$ to be unknown random variables, generated by their
own models conditioned on a small (non-extensive) number of
constraints. For the Poisson model in particular, this scenario is
relevant when we observe a network sampled from it, but have no direct
information on which values of $\bm\theta$ were used to generate it.

We begin with the microcanonical model, and assume that the degrees
are sampled uniformly at random, constrained only on their total sum,
$2E$. Since the number of different degree sequences is ${2E+N-1
  \choose 2E}$, the uniform probability is
\begin{equation}
  P(\bm k | E) = {2E+N-1 \choose 2E}^{-1},
\end{equation}
assuming $\sum_ik_i=2E$, otherwise $P(\bm k | E)=0$. We then assume that
the total sum is a sample from a Poisson distribution with mean
$\lambda$, $P(E|\lambda) = \lambda^E\e^{-\lambda}/E!$. This gives us
a total marginal distribution
\begin{align}
    P(\A|\lambda) &= \sum_{E'}\sum_{\bm k'}P(\A|\bm k')P(\bm k'|E')P(E'|\lambda)\\
    &=\frac{(2\lambda)^E\e^{-\lambda}\prod_ik_i!}{\prod_{i<j}A_{ij}!\prod_iA_{ii}!!}\frac{(N-1)!}{(2E+N-1)!},
\end{align}
which is nonzero for every $\A$.

Now turning to the Poisson model, we assume without loss of generality
the re-parametrization $\theta_i=\sqrt{2\lambda}\kappa_i$, with
$\sum_i\kappa_i=1$ and $\lambda \in [0,\infty]$, such that
Eq.~\ref{eq:poisson-sl} becomes
\begin{equation}
  P(\A|\bm\kappa,\lambda) = \frac{(2\lambda)^E\e^{-\lambda}\prod_i\kappa_i^{k_i}}{\prod_{i<j}A_{ij}!\prod_iA_{ii}!!}.
\end{equation}
We then assume that $\bm\kappa$ is sampled uniformly at random from the
simplex
\begin{equation}
  P(\bm\kappa) = (N-1)!\,\delta\left(\textstyle\sum_i\kappa_i - 1\right).\label{eq:kappa}
\end{equation}
Computing the marginal distribution, we obtain
\begin{align}
    P(\A|\lambda) &= \int P(\A|\bm\kappa,\lambda)P(\bm\kappa)\;\dd\bm\kappa\\
    &=\frac{(2\lambda)^E\e^{-\lambda}\prod_ik_i!}{\prod_{i<j}A_{ij}!\prod_iA_{ii}!!}\frac{(N-1)!}{(2E+N-1)!},
\end{align}
which is identical to the marginal obtained with the microcanonical model.

The above equivalence means that these two distinct generative
processes, involving either the configuration model or the Poisson
model, yield exactly the same marginal distribution over multigraphs. A
direct consequence of this is that, when all we observe is a single
network $\A$, there is no information contained in it that allows us to
determine whether it came from one of the two models. In statistical
terminology, we say these models are not identifiable.

Combining all of the above, we have that the following generative
processes are fully identical:
\begin{enumerate}
    \item \textbf{Poisson model:}
      \begin{enumerate}
        \item The relative fugacities $\bm\kappa$ are sampled uniformly
              at random from Eq.~\ref{eq:kappa}.
        \item Given the expected number of edges $\lambda$, the
              fugacities are given by
              $\theta_i=\sqrt{2\lambda}\kappa_i$, and the network is
              sampled from the Poisson model of Eq.~\ref{eq:poisson-sl}.
      \end{enumerate}
    \item \textbf{Sequential edge dropping:}
      \begin{enumerate}
        \item The total number of edges is sampled from a Poisson distribution with mean $\lambda$.
        \item The relative fugacities $\bm\kappa$ are sampled uniformly
              at random from Eq.~\ref{eq:kappa}.
        \item The network is sampled from the edge dropping model of
              Eq.~\ref{eq:edge_dropping}, with probabilities
              $q_{ij}\propto\kappa_i\kappa_j$ (and allowing for self-loops).
      \end{enumerate}
    \item \textbf{Configuration model:}
      \begin{enumerate}
        \item The total number of edges is sampled from a Poisson distribution with mean $\lambda$.
        \item The degrees are sampled uniformly at random from the set
          that preserves the total number of edges.
        \item The half-edges are paired uniformly at random.
      \end{enumerate}
\end{enumerate}

The above serves to demonstrate that we can arrive at the Poisson model
from several simple intuitive assumptions about the network formation
mechanism. These are all ``null'' models of network formation, since
they are not meant to realistically explain how networks in the real
world are formed, instead they contain only the smallest set of
ingredients necessary for a particular pattern --- in this case the
expected degrees of the nodes. What they all have in common is that,
during the network formation, potential multiple edges are treated as
individual elements, which is what lies behind the eventual equivalence
with the Poisson model.

\bibliography{bib}

\end{document}